\documentclass[10pt]{article}
\usepackage{graphicx}
\usepackage{amsmath}
\usepackage{amssymb}
\usepackage{caption2}
\begin{document}

\begin{center}
{\large {\bf \sc{The $\Lambda$-type P-wave  bottom baryon states via the QCD sum rules }}} \\[2mm]
Qi Xin ${}^{*\dagger}$, Zhi-Gang  Wang ${}^*$ \footnote{E-mail: zgwang@aliyun.com.  },
Fei Lu${}^*$

Department of Physics, North China Electric Power University, Baoding 071003, P. R. China ${}^*$\\
School of Nuclear Science and Engineering, North China Electric Power University, Beijing 102206, P. R. China ${}^\dagger$
\end{center}

\begin{abstract}
Our study focuses on the $\Lambda$-type P-wave  bottom  baryon states with the spin-parity $J^P=\frac{1}{2}^-$, $\frac{3}{2}^-$. We introduce an explicit P-wave between the two light quarks in the interpolating  currents
(and the two light quarks are antisymmetric in flavor space, therefore leads to the name $\Lambda$-type baryon) to investigate the $\Lambda_b$ and $\Xi_b$ states within the framework of  the full QCD sum rules.  The predicted masses show that the $\Xi_b(6087)$ and $\Xi_b(6095/6100)$ could to be the P-wave bottom-strange  baryon states with the spin-parity $J^P=\frac{1}{2}^-$ and $\frac{3}{2}^-$, respectively, meanwhile, the $\Lambda_b(5912)$ and $\Lambda_b(5920)$ could be the P-wave bottom baryon states with the spin-parity $J^P=\frac{1}{2}^-$ and $\frac{3}{2}^-$, respectively. The $\Lambda_b(5920)$ and $\Xi_b(6095/6100)$ maybe   have two remarkable under-structures or Fock components at least.
\end{abstract}



\section{Introduction}
In the past years, several excited bottom baryon states $\Lambda_b$ \cite{LHCb-6072,LHCb-6146-6152,LHCb-5912-5920} and excited strange-bottom baryon states $\Xi_b$ \cite{D0-5774,CDF-5793,CDF-5788,LHCb-6227-1,LHCb-6227-2,CMS-6100,LHCb-6327-6333} have been observed. In 2012, the LHCb collaboration observed  two narrow states $\Lambda_b(5912)^0$ and $\Lambda_b(5920)^0$ in the $\Lambda_b^0 \pi^+\pi^-$ invariant mass spectrum \cite{LHCb-5912-5920}, the masses were measured to be,
\begin{flalign}
& M_{\Lambda_b(5912)} = 5911.97\pm0.12 \pm0.02 \pm0.66 \mbox{ MeV}\,,\nonumber \\
& M_{\Lambda_b(5920)} = 5919.77 \pm0.08 \pm0.02 \pm0.66 \mbox{ MeV}\,.
\end{flalign}
In 2021, the CMS collaboration discovered the  $\Xi_b(6100)^-$ in the $\Xi_b^- \pi^+\pi^-$ invariant mass spectrum with the spin-parity $J^P=\frac{3}{2}^-$ \cite{CMS-6100}, the measured mass is
\begin{flalign}
&M_{\Xi_b(6100)} = 6100.3\pm0.2 \pm0.1 \pm0.6 \mbox{ MeV}\, .
\end{flalign}
Sometime ago, the LHCb collaboration confirmed the $\Xi_b(6100)^-$ in the decay mode $\Xi_b^{*0}\pi^-$, and observed the bottom baryon states $\Xi_b(6087)^0$ and  $\Xi_b(6095)^0$ in the decay modes $\Xi_b^{\prime-} \pi^+$ and $\Xi_b^{*-} \pi^+$, respectively  \cite{LHCb-6087-6095,LHCb-6087-6095-GKA}, the masses and widths are determined to be,
\begin{flalign}
& M_{\Xi_b(6100)} = 6099.74\pm0.11\pm0.02\pm0.6 \mbox{ MeV}\,,\,\,\Gamma_{\Xi_b(6100)} = 0.94\pm0.30\pm0.08 \mbox{ MeV} \, ,\nonumber \\
& M_{\Xi_b(6087)} = 6087.24\pm0.20 \pm0.06 \pm0.5 \mbox{ MeV}\,,\,\,\Gamma_{\Xi_b(6087)} = 2.43 \pm0.51 \pm0.10 \mbox{ MeV} \, ,\nonumber \\
& M_{\Xi_b(6095)} = 6095.26 \pm0.15 \pm0.03 \pm0.5 \mbox{ MeV}\,,\,\,\Gamma_{\Xi_b(6095)} = 0.50 \pm0.33 \pm0.11\mbox{ MeV} \, .
\end{flalign}
We can tentatively assign the $\Xi_b(6095)^0$ and $\Xi_b(6100)^-$ to be the isospin doublet by considering the mass difference and quark constituents.

Experimental discoveries of those single bottom baryon states have heightened the interest of theoretical researches. It is necessary to identify the quantum numbers of those states and explore their internal structures. The bottom baryon states have been investigated by many theoretical approaches, including the elementary emission model and ${}^3P_0$ model (the quark pair creation model) \cite{Santopinto-Xib} (\cite{LX-Xib}), the chiral quark model \cite{ZXH-Xib,LQF-Lambdab}, the constituent quark model \cite{LQF-Sigmab-Xib,CQM-ChenB,CQM-Kakadiya}, the QCD-motivated relativistic quark model \cite{Ebert-heavy-baryon}, the light-cone QCD sum rule \cite{Aliev-Sigmab}, the QCD sum rules \cite{WZG-WHJ-2S,HMQ-QCDSR,WZG-32-EPJC,ZhangJR,WZG-Negative-EPJA,Azizi-Lambdab,
Azizi-Xib,YGL-Xib-Lambdab,WZG-Omegab-1,ChenHX}, the lattice QCD \cite{Burch-lattice}, etc. Especially, the $\Lambda_b(5912)$ and $\Lambda_b(5920)$ have been investigated by the flux tube model \cite{CB-5912-5920}, the QCD sum rules combined with  the  heavy quark effective theory \cite{CHX-5912-5920}, the effective hadronic model with respecting the chiral symmetry and heavy-quark spin-flavor symmetry \cite{Kawakami-5912-5920},  the relativized quark model \cite{LZY-lambdaQ}, etc. Those studies show that the $\Lambda_b(5912)$ and $\Lambda_b(5920)$ can be accommodated in the 1P states with the $J^P=\frac{1}{2}^-$ and $\frac{3}{2}^-$, respectively. Additionally, the $\Xi_b(6100)$ can be taken as a good candidate for the 1P bottom state with the $J^P=\frac{3}{2}^-$ by the ${}^3P_0$ model \cite{LQF-6100}, the QCD sum rules combined with the heavy quark effective theory \cite{CHX-6100},
the relativized quark model \cite {YGL-XiQ}, etc.

In our previous works, we have used the full QCD sum rules to  investigate the heavy baryon states systematically. We calculated the masses of the S-wave, P-wave and D-wave charmed baryon candidates  $\Omega_c(3000)$, $\Omega_c(3050)$, $\Omega_c(3066)$, $\Omega_c(3090)$, $\Omega_c(3119)$ \cite{WZG-Omegac-1,WZG-Omegac-2}, $\Omega_c(3327)$ \cite{WZG-Omegac-D}, $\Lambda_c(2625)$ \cite{WZG-Lambdac-Xic-P}, $\Xi_c(2815)$ \cite{WZG-Lambdac-Xic-P}, $\Lambda_c(2860)$, $\Lambda_c(2880)$,  $\Xi_c(3055)$, $\Xi_c(3080)$ \cite{WZG-Lambdac-Xic-D}, and acquired satisfactory results which are consistent with the experimental data and serve as guides for the future experimental measurements. On the bottom sector, our calculations also lead to satisfactory assignments for the  P-wave candidates $\Omega_b(6316)$, $\Omega_b(6330)$, $\Omega_b(6340)$, $\Omega_b(6350)$ \cite{WZG-Omegab-1} and D-wave candidates  $\Lambda_b(6146)$, $\Lambda_b(6152)$, $\Xi_b(6327)$, $\Xi_b(6333)$ \cite{YGL-Xib-Lambdab}.

In the constituent quark models, the $\Xi_b$ states have three valence quarks $q\,(u,d)$, $s$ and $b$. Introducing  the relative P-wave between the $q$ and $s$ in the diquarks, we obtain the $\Lambda$-type $\Xi_b$ baryon states with the spin-parity  $J^P=\frac{1}{2}^-$, $\frac{3}{2}^-$. We can choose either partial  derivative $\partial_\mu$ or covariant derivative $D_\mu=\partial_\mu-ig_sG_\mu$ to embody the net effects of the relative P-wave. It is interesting to choose both partial derivatives $\partial_\mu$  and covariant derivatives $D_\mu$ in constructing the P-wave states, and examine the different outcomes, just like in our previous works on the D-wave charmed baryon states \cite{WZG-Omegac-D}. In this work, we explore the P-wave $\Lambda$-type bottom baryon states in details with the QCD sum rules, and examine the sub-structures of the new bottom baryon states $\Lambda_b(5912)$, $\Lambda_b(5920)$,  $\Xi_b(6087)$, $\Xi_b(6095)$ and $\Xi_b(6100)$ so as to diagnose their nature, because their properties are not well understood yet.

The paper is structured as follows: the P-wave bottom baryon states are studied via the QCD sum rules in section 2; the numerical results and discussions are displayed in section 3; in section 4, the relevant conclusions are obtained.

\section{QCD sum rules for the P-wave bottom baryons states}
Firstly, we write down the two-point correlation functions $\Pi(p)$ and $\Pi_{\mu\nu}(p)$,
\begin{eqnarray}
\Pi(p)&=&i\int d^4x e^{ip \cdot x} \langle0|T\left\{J/\eta(x) \bar{ J}/\bar{\eta}(0)\right
\}|0\rangle \, , \nonumber\\
\Pi_{\mu\nu}(p)&=&i\int d^4x e^{ip \cdot x} \langle0|T\left\{J/\eta_\mu(x)\bar{J}/\bar{\eta}_\nu(0)\right
\}|0\rangle \, ,
\end{eqnarray}
where
\begin{eqnarray}
J(x)&=&J^{\Lambda_b}(x)\, , \,\,  J^{\Xi_b}(x)\, ,\nonumber \\
\eta(x)&=& \eta^{\Lambda_b}(x)\, , \,\,\eta^{\Xi_b}(x)\, ,\nonumber \\
J_\mu(x)&=&J_{1,\mu}^{\Lambda_b}(x)\, ,\,\,J_{2,\mu}^{\Lambda_b}(x)\, ,\,\,  J_{1,\mu}^{\Xi_b}(x)\, ,\,\,J_{2,\mu}^{\Xi_b}(x)\, , \nonumber \\
\eta_\mu(x)&=& \eta_{1,\mu}^{\Lambda_b}(x)\, , \,\, \eta_{2,\mu}^{\Lambda_b}(x)\, ,\, \,
\eta_{1,\mu}^{\Xi_b}(x)\, ,\,\,\eta_{2,\mu}^{\Xi_b}(x)\, ,
\end{eqnarray}

\begin{eqnarray}
J^{\Lambda_b}(x) &=&\varepsilon^{ijk} \left[ \partial^\mu u^T_i(x) C\gamma^\nu d_j(x)- u^T_i(x) C\gamma^\nu \partial^{\mu}d_j(x)\right]\sigma_{\mu\nu}\,b_k(x) \, ,\nonumber \\
J_{1,\mu}^{\Lambda_b}(x)&=&\varepsilon^{ijk} \left[ \partial^\alpha u^T_i(x) C\gamma^\beta d_j(x)-u^T_i(x) C\gamma^\beta \partial^{\alpha}d_j(x)\right]\left(\widetilde{g}_{\mu\alpha}\gamma_\beta-\widetilde{g}_{\mu\beta}\gamma_\alpha \right)i\gamma_5 b_k(x) \, ,\nonumber \\
J_{2,\mu}^{\Lambda_b}(x)&=&\varepsilon^{ijk} \left[ \partial^\alpha u^T_i(x) C\gamma^\beta d_j(x)- u^T_i(x) C\gamma^\beta \partial^{\alpha}d_j(x)\right]
\left(g_{\mu\alpha}\gamma_\beta+g_{\mu\beta}\gamma_\alpha-\frac{1}{2}g_{\alpha\beta}\gamma_\mu \right)i\gamma_5 b_k(x)\, , \nonumber \\
\end{eqnarray}

\begin{eqnarray}
\eta^{\Lambda_b}(x) &=&\varepsilon^{ijk} \left[ D^\mu u^T_i(x) C\gamma^\nu d_j(x)- u_i(x) C\gamma^\nu D^{\mu}d_j(x)\right]\sigma_{\mu\nu}\,b_k(x) \, ,\nonumber \\
\eta_{1,\mu}^{\Lambda_b}(x)&=&\varepsilon^{ijk} \left[ D^\alpha u^T_i(x) C\gamma^\beta d_j(x)-u^T_i(x) C\gamma^\beta D^{\alpha}d_j(x)\right]\left(\widetilde{g}_{\mu\alpha}\gamma_\beta-\widetilde{g}_{\mu\beta}\gamma_\alpha \right)i\gamma_5 b_k(x) \, ,\nonumber \\
\eta_{2,\mu}^{\Lambda_b}(x) &=&\varepsilon^{ijk} \left[ D^\alpha u^T_i(x) C\gamma^\beta d_j(x)- u^T_i(x) C\gamma^\beta D^{\alpha}d_j(x)\right]
\left(g_{\mu\alpha}\gamma_\beta+g_{\mu\beta}\gamma_\alpha-\frac{1}{2}g_{\alpha\beta}\gamma_\mu \right)i\gamma_5 b_k(x) \, , \nonumber \\
\end{eqnarray}

\begin{eqnarray}
J^{\Xi_b}(x) &=&\varepsilon^{ijk} \left[ \partial^\mu q^T_i(x) C\gamma^\nu s_j(x)- q^T_i(x) C\gamma^\nu \partial^{\mu}s_j(x)\right]\sigma_{\mu\nu}\,b_k(x) \, ,\nonumber \\
J_{1,\mu}^{\Xi_b}(x)&=&\varepsilon^{ijk} \left[ \partial^\alpha q^T_i(x) C\gamma^\beta s_j(x)-q^T_i(x) C\gamma^\beta \partial^{\alpha}s_j(x)\right]\left(\widetilde{g}_{\mu\alpha}\gamma_\beta-\widetilde{g}_{\mu\beta}\gamma_\alpha \right)i\gamma_5 b_k(x) \, ,\nonumber \\
J_{2,\mu}^{\Xi_b}(x)&=&\varepsilon^{ijk} \left[ \partial^\alpha q^T_i(x) C\gamma^\beta s_j(x)- q^T_i(x) C\gamma^\beta \partial^{\alpha}s_j(x)\right]
\left(g_{\mu\alpha}\gamma_\beta+g_{\mu\beta}\gamma_\alpha-\frac{1}{2}g_{\alpha\beta}\gamma_\mu \right)i\gamma_5 b_k(x)\, , \nonumber \\
\end{eqnarray}

\begin{eqnarray}
\eta^{\Xi_b}(x) &=&\varepsilon^{ijk} \left[ D^\mu q^T_i(x) C\gamma^\nu s_j(x)- q^T_i(x) C\gamma^\nu D^{\mu}s_j(x)\right]\sigma_{\mu\nu}\,b_k(x) \, ,\nonumber \\
\eta_{1,\mu}^{\Xi_b}(x)&=&\varepsilon^{ijk} \left[ D^\alpha q^T_i(x) C\gamma^\beta s_j(x)-q^T_i(x) C\gamma^\beta D^{\alpha}s_j(x)\right]\left(\widetilde{g}_{\mu\alpha}\gamma_\beta-\widetilde{g}_{\mu\beta}\gamma_\alpha \right)i\gamma_5 b_k(x) \, ,\nonumber \\
\eta_{2,\mu}^{\Xi_b}(x)&=&\varepsilon^{ijk} \left[D^\alpha q^T_i(x) C\gamma^\beta s_j(x)- q^T_i(x) C\gamma^\beta D^{\alpha}s_j(x)\right]
\left(g_{\mu\alpha}\gamma_\beta+g_{\mu\beta}\gamma_\alpha-\frac{1}{2}g_{\alpha\beta}\gamma_\mu \right)i\gamma_5 b_k(x)\, , \nonumber \\
\end{eqnarray}
 $q=u$ or $d$, the $i$, $j$, $k$ are color indexes, the $C$ is the charge conjugation matrix, the tensor structure $\widetilde{g}_{\mu\alpha}=g_{\mu\alpha}-\frac{1}{4}\gamma_\mu\gamma_\alpha$. We choose both  the partial derivatives $\partial_\mu$  and covariant derivatives $D_\mu$ to construct the currents, the $J/\eta(x)$ and $J/\eta_\mu(x)$ interpolate the P-wave baryon states with the spin-parity $J^P={\frac{1}{2}}^-$ and ${\frac{3}{2}}^-$, respectively.  The currents with  covariant derivatives   are gauge covariant/invariant, but disfavors  interpreting  the covariant derivatives as  angular momentum in the non-relativistic limit, $D \to \vec{p}+g_s\vec{G}$. On the other hand, the currents with partial derivatives $\partial_\mu$ are not gauge covariant, but favors  interpreting  the partial derivatives as  angular momentum in the non-relativistic limit, $\partial \to \vec{p}$. In the quantum field theory, we construct gauge invariant  currents with the same quantum numbers as the hadrons to interpolate them, that is enough. In this sense, the gauge invariant  currents are physical and are preferred.

 The  diquarks  $\varepsilon^{ijk}  q^T_i(x) C\gamma_\alpha \stackrel{\leftrightarrow}{\partial}_\beta q^\prime_j(x)$ and $\varepsilon^{ijk}  q^T_i(x) C\gamma_\alpha \stackrel{\leftrightarrow}{D}_\beta q^\prime_j(x)$ have  two Lorentz indexes $\alpha$ and $\beta$, where $q\neq q^\prime$,
 $\stackrel{\leftrightarrow}{\partial}_\beta=\stackrel{\rightarrow}{\partial}_\beta
 -\stackrel{\leftarrow}{\partial}_\beta$ and $\stackrel{\leftrightarrow}{D}_\beta=\stackrel{\rightarrow}{D}_\beta-\stackrel{\leftarrow}{D}_\beta$.
The structures $C\gamma_\alpha \stackrel{\leftrightarrow}{\partial}_\beta $ and $C\gamma_\alpha \stackrel{\leftrightarrow}{D}_\beta $ are antisymmetric, therefore the currents $J/\eta(x)$  and $J/\eta_\mu(x)$ are  referred to as the $\Lambda$-type currents.
 The Dirac matrixes $\widetilde{g}_{\mu\alpha}\gamma_\beta-\widetilde{g}_{\mu\beta}\gamma_\alpha  $ and $g_{\mu\alpha}\gamma_\beta+g_{\mu\beta}\gamma_\alpha-\frac{1}{2}g_{\alpha\beta}\gamma_\mu$ are anti-symmetric and symmetric respectively when interchanging
  the indexes $\alpha$ and $\beta$, which are contracted with the corresponding indexes in the diquarks,  therefore  the diquarks in the currents $J/\eta_{1,\mu}(x)$ and $J/\eta_{2,\mu}(x)$ have the spins 1 and 2, respectively.

The currents $J/\eta(0)$ and $J/\eta_\mu(0)$ couple potentially to the $J^P={\frac{1}{2}}^\mp$ and ${\frac{1}{2}}^\pm$, ${\frac{3}{2}}^\mp$ bottom  baryon states $B_{\frac{1}{2}}^\mp$ and $B_{\frac{1}{2}}^\pm$, $B_{\frac{3}{2}}^\mp$, respectively,
\begin{eqnarray}\label{J-lamda-1}
\langle 0| J/\eta (0)|B_{\frac{1}{2}}^{-}(p)\rangle &=&\lambda^{-}_{\frac{1}{2}} U^{-}(p,s) \, ,\nonumber  \\
\langle 0| J/\eta (0)|B_{\frac{1}{2}}^{+}(p)\rangle &=&\lambda^{+}_{\frac{1}{2}}i\gamma_5 U^{+}(p,s) \, , \nonumber \\
\langle 0| J/\eta_{\mu} (0)|B_{\frac{3}{2}}^{-}(p)\rangle &=&\lambda^{-}_{\frac{3}{2}} U^{-}_\mu(p,s) \, , \nonumber \\
\langle 0| J/\eta_{\mu} (0)|B_{\frac{3}{2}}^{+}(p)\rangle &=&\lambda^{+}_{\frac{3}{2}}i\gamma_5 U^{+}_{\mu}(p,s) \, ,
\end{eqnarray}

\begin{eqnarray}\label{J-lamda-2}
\langle 0| J/\eta_{\mu} (0)|B_{\frac{1}{2}}^{+}(p)\rangle &=&\lambda^{+}_{\frac{1}{2}} p_\mu U^{+}(p,s) \, , \nonumber \\
\langle 0| J/\eta_{\mu} (0)|B_{\frac{1}{2}}^{-}(p)\rangle &=&\lambda^{-}_{\frac{1}{2}}i\gamma_5 p_\mu U^{-}(p,s) \, ,
\end{eqnarray}
where the $U^\pm(p,s)$ and $U^{\pm}_\mu(p,s)$  are the Dirac and Rarita-Schwinger spinors, respectively, the
$\lambda^{\pm}_{\frac{1}{2}}$ and $\lambda^{\pm}_{\frac{3}{2}}$ are the corresponding pole residues \cite{Chung82-1,Chung82-2,Oka96,ZGW-Pc4312-penta,ZGW-DSigmac-penta-mole,Wang1508-EPJC,WangHuang1508-1,WangHuang1508-2,WangHuang1508-3}.
The Rarita-Schwinger spinors $U^{\pm}_\mu(p,s)$ satisfy the relations $\gamma^\mu U^{\pm}_\mu(p,s)=0$, which correspond to the relations $\gamma^\mu J_\mu(x)=\gamma^\mu \eta_\mu(x)=0$. In general, the currents $J/\eta_\mu(x)$ are not necessary to satisfy such relations, however, in the present case, they happen to have such relations. So the Eq.\eqref{J-lamda-2} should be modified to be
\begin{eqnarray}\label{J-lamda-3}
\langle 0| J/\eta_{\mu} (0)|B_{\frac{1}{2}}^{+}(p)\rangle &=&\lambda^{+}_{\frac{1}{2}}\left(\gamma_\mu-4\frac{p_\mu}{M_{+}}\right) U^{+}(p,s) \, , \nonumber \\
\langle 0| J/\eta_{\mu} (0)|B_{\frac{1}{2}}^{-}(p)\rangle &=&\lambda^{-}_{\frac{1}{2}}i\gamma_5 \left(\gamma_\mu-4\frac{p_\mu}{M_{-}}\right) U^{-}(p,s) \, ,
\end{eqnarray}

At the hadron side of the correlation functions $\Pi(p)$ and $\Pi_{\mu\nu}(p)$, we isolate the  ground state contributions from the spin-parity $J^P={\frac{1}{2}}^\mp$ and ${\frac{3}{2}}^\mp$ baryon states according to the current-hadron couplings shown in Eqs.\eqref{J-lamda-1}-\eqref{J-lamda-3}, and get the hadronic representation \cite{QCDSR-SVZ79,QCDSR-Reinders85},
\begin{eqnarray}
\Pi(p) & = & {\lambda^{-}_{\frac{1}{2}}}^2  {\!\not\!{p}+ M_{-} \over M_{-}^{2}-p^{2}  } +  {\lambda^{+}_{\frac{1}{2}}}^2  {\!\not\!{p}- M_{+} \over M_{+}^{2}-p^{2}  } +\cdots  \, ,\nonumber\\
  &=&\Pi_{\frac{1}{2}}^1(p^2)\!\not\!{p}+\Pi_{\frac{1}{2}}^0(p^2)\, ,
\end{eqnarray}

\begin{eqnarray}
\Pi_{\mu\nu}(p) & = & {\lambda^{-}_{\frac{3}{2}}}^2  {\!\not\!{p}+ M_{-} \over M_{-}^{2}-p^{2}  } \left(- g_{\mu\nu}+\frac{\gamma_\mu\gamma_\nu}{3}+\frac{2p_\mu p_\nu}{3p^2}-\frac{p_\mu\gamma_\nu-p_\nu \gamma_\mu}{3\sqrt{p^2}}\right)\nonumber\\
&&+ {\lambda^{+}_{\frac{3}{2}}}^2  {\!\not\!{p}- M_{+} \over M_{+}^{2}-p^{2}  } \left(- g_{\mu\nu}+\frac{\gamma_\mu\gamma_\nu}{3}+\frac{2p_\mu p_\nu}{3p^2}-\frac{p_\mu\gamma_\nu-p_\nu \gamma_\mu}{3\sqrt{p^2}}\right)   \nonumber\\
&&+{\lambda^{+}_{\frac{1}{2}}}^2 \left(\gamma_\mu-4\frac{p_\mu}{M_{+}}\right) {\!\not\!{p}+ M_{+} \over M_{+}^{2}-p^{2}  }\left(\gamma_\nu-4\frac{p_\nu}{M_{+}}\right) \nonumber\\
&&+{\lambda^{-}_{\frac{1}{2}}}^2 \left(\gamma_\mu+4\frac{p_\mu}{M_{-}}\right) {\!\not\!{p}- M_{-} \over M_{-}^{2}-p^{2}  }\left(\gamma_\nu+4\frac{p_\nu}{M_{-}}\right)+\cdots \nonumber\\
&=&-\Pi_{\frac{3}{2}}^1(p^2)\!\not\!{p}\,g_{\mu\nu}-\Pi_{\frac{3}{2}}^0(p^2)\,g_{\mu\nu}+\cdots\, ,
\end{eqnarray}
we choose the components $\Pi_{\frac{1}{2}}^1(p^2)$, $\Pi_{\frac{1}{2}}^0(p^2)$, $\Pi_{\frac{3}{2}}^1(p^2)$ and $\Pi_{\frac{3}{2}}^0(p^2)$ to explore the spin-parity $J^{P}={\frac{1}{2}}^-$ and ${\frac{3}{2}}^-$ states, respectively, so as to avoid possible contaminations.

At the QCD side, we take the following full light-quark propagator $S_{ij}(x)$ and full heavy-quark  propagator $B_{ij}(x)$ when calculating the operator product expansion for the correlation functions $\Pi(p)$ and $\Pi_{\mu\nu}(p)$ \cite{QCDSR-Reinders85,Pascual-1984,WZG-HT-PRD},
\begin{eqnarray}\label{S-propagt}
S_{ij}(x)&=& \frac{i\delta_{ij}\!\not\!{x}}{ 2\pi^2x^4}
-\frac{\delta_{ij}m_q}{4\pi^2x^2}-\frac{\delta_{ij}\langle
\bar{q}q\rangle}{12} +\frac{i\delta_{ij}\!\not\!{x}m_q
\langle\bar{q}q\rangle}{48}-\frac{\delta_{ij}x^2\langle \bar{q}g_s\sigma Gq\rangle}{192}\nonumber\\
&&+\frac{i\delta_{ij}x^2\!\not\!{x} m_q\langle \bar{q}g_s\sigma
 Gq\rangle }{1152} -\frac{ig_s G^{a}_{\alpha\beta}t^a_{ij}(\!\not\!{x}
\sigma^{\alpha\beta}+\sigma^{\alpha\beta} \!\not\!{x})}{32\pi^2x^2} -\frac{1}{8}\langle\bar{q}_j\sigma^{\mu\nu}q_i \rangle \sigma_{\mu\nu} +\cdots \, ,
\end{eqnarray}
where $q=u$, $d$ or $s$, and
\begin{eqnarray}
B_{ij}(x)&=&\frac{i}{(2\pi)^4}\int d^4k e^{-ik \cdot x} \left\{
\frac{\delta_{ij}}{\!\not\!{k}-m_b}
-\frac{g_sG^n_{\alpha\beta}t^n_{ij}}{4}\frac{\sigma^{\alpha\beta}(\!\not\!{k}+m_b)+(\!\not\!{k}+m_b)
\sigma^{\alpha\beta}}{(k^2-m_b^2)^2}\right.\nonumber\\
&&\left. -\frac{g_s^2 (t^at^b)_{ij} G^a_{\alpha\beta}G^b_{\mu\nu}(f^{\alpha\beta\mu\nu}+f^{\alpha\mu\beta\nu}+f^{\alpha\mu\nu\beta}) }{4(k^2-m_b^2)^5}+\cdots\right\} \, ,
\end{eqnarray}
\begin{eqnarray}
f^{\alpha\beta\mu\nu}&=&(\!\not\!{k}+m_b)\gamma^\alpha(\!\not\!{k}+m_b)\gamma^\beta(\!\not\!{k}+m_b)
\gamma^\mu(\!\not\!{k}+m_b)\gamma^\nu(\!\not\!{k}+m_b)\, ,
\end{eqnarray}
for the technical details, one can consult Ref.\cite{WZG-HT-PRD}.

Similar to previous works \cite{ZGW-Pc4312-penta,ZGW-DSigmac-penta-mole,Wang1508-EPJC,WangHuang1508-1,WangHuang1508-2,WangHuang1508-3}, we select the components associated with the structures  $\!\not\!{p}$, $1$,  $\!\not\!{p} g_{\mu\nu}$ and $g_{\mu\nu}$ in the correlation functions  $\Pi(p)$ and $\Pi_{\mu\nu}(p)$ to investigate the baryon states with the spin-parity $J^P=\frac{1}{2}^\mp$ and $\frac{3}{2}^\mp$, respectively. Then we obtain the spectral densities at the hadronic side through dispersion relation,
\begin{eqnarray}
\frac{{\rm Im}\Pi_{j}^1(s)}{\pi}&=& {\lambda^{-}_{j}}^2 \delta\left(s-M_{-}^2\right)+{\lambda^{+}_{j}}^2 \delta\left(s-M_{+}^2\right) =\, \rho^1_{j,H}(s) \, , \\
\frac{{\rm Im}\Pi^0_{j}(s)}{\pi}&=&M_{-}{\lambda^{-}_{j}}^2 \delta\left(s-M_{-}^2\right)-M_{+}{\lambda^{+}_{j}}^2 \delta\left(s-M_{+}^2\right)
=\rho^0_{j,H}(s) \, ,
\end{eqnarray}
where $j=\frac{1}{2}$, $\frac{3}{2}$,  the subscript $H$ represents the hadron side,
then we introduce the  weight functions $\sqrt{s}\exp\left(-\frac{s}{T^2}\right)$ and $\exp\left(-\frac{s}{T^2}\right)$ to obtain the QCD sum rules at the phenomenological side,
\begin{eqnarray}\label{QCDSR-M}
\int_{m_b^2}^{s_0}ds \left[\sqrt{s}\rho^1_{j,H}(s)+\rho^0_{j,H}(s)\right]\exp\left( -\frac{s}{T^2}\right)
&=&2M_{-}{\lambda^{-}_{j}}^2\exp\left( -\frac{M_{-}^2}{T^2}\right) \, ,
\end{eqnarray}
where the $s_0$ are the continuum threshold parameters and the $T^2$ are the Borel parameters.
We separate the  contributions  of the negative-parity baryon states from that of the positive-parity baryon states unambiguously. In Eq.\eqref{QCDSR-M}, the threshold is taken as $m_b^2$ rather than $(m_b+m_q+m_{q^\prime})^2$, which is consistent with the light-quark propagator shown in Eq.\eqref{S-propagt}, where the small light quark mass is taken as perturbative correction and does not modify dispersion relation.

We differentiate   Eq.\eqref{QCDSR-M} with respect to  $\tau=\frac{1}{T^2}$, then eliminate the pole residues $\lambda^{-}_{j}$ with $j=\frac{1}{2}$ and $\frac{3}{2}$ to obtain the QCD sum rules for the masses of the  P-wave  baryons states,
\begin{eqnarray}
M^2_{-} &=& \frac{-\frac{d}{d \tau}\int_{m_b^2}^{s_0}ds \,\left[\sqrt{s}\,\rho^1_{QCD}(s)+\,\rho^0_{QCD}(s)\right]\exp\left(- \tau s\right)}{\int_{m_b^2}^{s_0}ds \left[\sqrt{s}\,\rho_{QCD}^1(s)+\,\rho^0_{QCD}(s)\right]\exp\left( -\tau s\right)}\, ,
\end{eqnarray}
where the spectral densities $\rho_{QCD}^1(s)=\rho_{j,QCD}^1(s)$ and $\rho^0_{QCD}(s)=\rho^0_{j,QCD}(s)$, the explicit expressions are given in the Appendix.

\section{Numerical results and discussions}
We choose the standard values of the vacuum condensates
$\langle\bar{q}q \rangle=-(0.24\pm 0.01\, \rm{GeV})^3$,  $\langle\bar{s}s \rangle=(0.8\pm0.1)\langle\bar{q}q \rangle$, $\langle\bar{q}g_s\sigma G q \rangle=m_0^2\langle \bar{q}q \rangle$,
 $\langle\bar{s}g_s\sigma G s \rangle=m_0^2\langle \bar{s}s \rangle$,
$m_0^2=(0.8 \pm 0.1)\,\rm{GeV}^2$, $\langle \frac{\alpha_s
GG}{\pi}\rangle=(0.33\,\rm{GeV})^4 $    at the energy scale  $\mu=1\, \rm{GeV}$
\cite{QCDSR-SVZ79,QCDSR-Reinders85,QCDSR-Colangelo-Review}, the $\overline{MS}$ masses $m_{b}(m_b)=(4.18\pm0.03)\,\rm{GeV}$ and $m_s(\mu=2\,\rm{GeV})=(0.095\pm0.005)\,\rm{GeV}$
from the Particle Data Group \cite{PDG-2020}. We want to extract the masses of the P-wave baryons states at the best energy scales $\mu$ of the QCD spectral densities, as the input parameters  evolve with the energy scale $\mu$ according to the re-normalization  group equation,
\begin{eqnarray}
 \langle\bar{q}q \rangle(\mu)&=&\langle\bar{q}q\rangle({\rm 1 GeV})\left[\frac{\alpha_{s}({\rm 1 GeV})}{\alpha_{s}(\mu)}\right]^{\frac{12}{33-2n_f}}\, , \nonumber\\
 \langle\bar{s}s \rangle(\mu)&=&\langle\bar{s}s \rangle({\rm 1 GeV})\left[\frac{\alpha_{s}({\rm 1 GeV})}{\alpha_{s}(\mu)}\right]^{\frac{12}{33-2n_f}}\, , \nonumber\\
 \langle\bar{q}g_s \sigma Gq \rangle(\mu)&=&\langle\bar{q}g_s \sigma Gq \rangle({\rm 1 GeV})\left[\frac{\alpha_{s}({\rm 1 GeV})}{\alpha_{s}(\mu)}\right]^{\frac{2}{33-2n_f}}\, ,\nonumber\\
  \langle\bar{s}g_s \sigma Gs \rangle(\mu)&=&\langle\bar{s}g_s \sigma Gs \rangle({\rm 1 GeV})\left[\frac{\alpha_{s}({\rm 1 GeV})}{\alpha_{s}(\mu)}\right]^{\frac{2}{33-2n_f}}\, ,\nonumber\\
m_b(\mu)&=&m_b(m_b)\left[\frac{\alpha_{s}(\mu)}{\alpha_{s}(m_b)}\right]^{\frac{12}{33-2n_f}} \, ,\nonumber\\
m_s(\mu)&=&m_s({\rm 2GeV} )\left[\frac{\alpha_{s}(\mu)}{\alpha_{s}({\rm 2GeV})}\right]^{\frac{12}{33-2n_f}}\, ,\nonumber\\
\alpha_s(\mu)&=&\frac{1}{b_0t}\left[1-\frac{b_1}{b_0^2}\frac{\log t}{t} +\frac{b_1^2(\log^2{t}-\log{t}-1)+b_0b_2}{b_0^4t^2}\right]\, ,
\end{eqnarray}
where $t=\log \frac{\mu^2}{\Lambda_{QCD}^2}$, $b_0=\frac{33-2n_f}{12\pi}$, $b_1=\frac{153-19n_f}{24\pi^2}$, $b_2=\frac{2857-\frac{5033}{9}n_f+\frac{325}{27}n_f^2}{128\pi^3}$,  $\Lambda_{QCD}=210\,\rm{MeV}$, $292\,\rm{MeV}$  and  $332\,\rm{MeV}$ for the flavors  $n_f=5$, $4$ and $3$, respectively  \cite{PDG-2020,Narison-mix},  we can take the flavor numbers  $n_f=5$ for the bottom baryon states.

We calculate the vacuum condensates in the operator product expansion up  to dimension 10, and study the P-wave bottom baryon states by considering  the light flavor $SU_f(3)$ breaking effects. We take the modified  energy scale formula $ \mu =\sqrt{M_{B}^2-{\mathbb{M}}_b^2}-k{\mathbb{M}}_s$, where the $k$  is the number of the $s$-quark in the currents \cite{WZG-Omegac-D,WZG-EPJC-bb,WZG-IJMPA-mole}, $M_B=M_{-}$, the ${\mathbb{M}}_b$ and ${\mathbb{M}}_s$ are the effective $b$-quark and $s$-quark masses, respectively. In order to ensure that the QCD spectral densities are taken at the best energy scales $\mu$, we choose the effective $b$-quark mass ${\mathbb{M}}_b=5.17\,\rm{GeV}$ and effective $s$-quark mass ${\mathbb{M}}_s=0.2\,\rm{GeV}$, which are fitted in the QCD sum rules for the tetraquark  states  \cite{WZG-EPJC-bb,WZG-IJMPA-mole}.

 We acquire the best energy scales $\mu$ and other parameters via trial and error, then obtain the  masses and pole residues of those P-wave bottom baryon states, the numerical values of the energy scales, continuum threshold parameters, Borel windows, pole contributions, perturbative contributions, masses  and  pole residues are shown in Tables \ref{Borel}-\ref{mass}. From the tables, as one can see,  the continuum threshold parameters and predicted baryon masses  have the relation $\sqrt{s_0}-M_{B}= 0.60\sim0.70\pm0.1\,\rm{GeV}$, which satisfies  our naive expectations of the mass gaps between the ground states and excited states. Furthermore,   the pole contributions are about $(40-65)\%$, and dominant contributions come from the perturbative terms, it is reasonable to extract the hadron masses. We choose the pole contributions about $(40-65)\%$, and the central values exceed $50\%$, just like what we have done in our previous works for other S-wave, P-wave and D-wave bottom baryon states \cite{WZG-WHJ-2S,YGL-Xib-Lambdab,WZG-Omegab-1}.

 If we prefer larger pole contributions, we should re-choose the parameters and Borel windows,  the resulting  energy scales, continuum threshold parameters, Borel windows, pole contributions, perturbative contributions, masses  and  pole residues are shown in Tables \ref{example-Borel}-\ref{example-mass}. From the Tables \ref{Borel}-\ref{example-mass}, we can see clearly that larger pole contributions lead to larger pole residues while the predicted masses are almost unchanged, moreover, we have to choose larger continuum threshold parameters, which maybe lead to some contaminations from the excited states or higher resonances and weaken the predictive power. Therefore we should perform a comprehensive  analysis for all the S-wave, P-wave and D-wave baryon states with the pole contributions larger than $50\%$ in a self-consistent way.  According to the huge load of work, we prefer to do it  in our next work.

In calculations, we find that the  differences between the central values of the baryon masses from the currents $J_{(\mu)}(x)$ and $\eta_{(\mu)} (x)$ are less than $0.02\,\rm{GeV}$ if choosing the same parameters. Slightly changing the Borel windows $T^2$ or continuum threshold parameters $s_0$  are sufficient to smear  the differences between the outcomes of the partial derivatives and covariant derivatives. From Tables \ref{Borel} and \ref{mass}, we happen to find, if we pick the same pole contributions, then the central values of the baryon masses remain essentially unchanged while that of the pole residues change slightly.
Another interesting thing is that  the contributions of the  perturbative  terms for the currents $\eta_{(\mu)} (x)$ with covariant derivatives are smaller than that from the currents $J_{(\mu)}(x)$ with partial derivatives.
The currents with  covariant derivatives   are gauge covariant/invairant, but disfavors  interpreting  the covariant derivatives as  angular momentum in the non-relativistic limit, $D \to \vec{p}+g_s\vec{G}$, while the currents with partial derivatives  are not gauge covariant, but favors  interpreting  the partial derivatives as  angular momentum in the non-relativistic limit, $\partial \to \vec{p}$. If only the baryon masses are concerned, we can choose either the currents $J_{(\mu)}(x)$ or $\eta_{(\mu)} (x)$.

\begin{table}
\begin{center}
\begin{tabular}{|c|c|c|c|c|c|c|c|c|}\hline\hline
\rm{Currents}   &$J^P$ &$\mu$      &$T^2(\rm{GeV}^2)$ &$\sqrt{s_0}(\rm GeV) $ &pole &$\rm{perturbative}$\\ \hline

$J^{\Lambda_b}$ &$\frac{1}{2}^-$   &$2.9$     &$3.6-4.0$      &$6.55\pm0.1$     &$(40-62)\%$  &$(90-94)\%$ \\

$J_{1,\mu}^{\Lambda_b}$   &$\frac{3}{2}^-$       &$2.9$                &$3.6-4.0$         &$6.55\pm0.1$               &$(41-63)\%$       &$(88-92)\%$                           \\

$J_{2,\mu}^{\Lambda_b}$   &$\frac{3}{2}^-$      &$2.9$             &$3.8-4.2$            &$6.60\pm0.1$               &$(43-64)\%$       &$(85-90)\%$ \\   \hline

$\eta^{\Lambda_b}$          &$\frac{1}{2}^-$   &$2.9$             &$3.7-4.1$            &$6.55\pm0.1$              &$(40-61)\%$        &$(85-89)\%$       \\

$\eta_{1,\mu}^{\Lambda_b}$   &$\frac{3}{2}^-$  &$2.9$            &$3.7-4.1$         &$6.55\pm0.1$               &$(40-61)\%$       &$(83-88)\%$                           \\

$\eta_{2,\mu}^{\Lambda_b}$   &$\frac{3}{2}^-$   &$2.9$           &$3.8-4.2$            &$6.60\pm0.1$               &$(42-61)\%$       &$(77-84)\%$ \\   \hline

$J^{\Xi_b}$              &$\frac{1}{2}^-$       &$3.0$             &$3.9-4.3$            &$6.70\pm0.1$               &$(42-64)\%$        &$(95-97)\%$                            \\

$J_{1,\mu}^{\Xi_b}$      &$\frac{3}{2}^-$       &$3.0$               &$3.9-4.3$            &$6.70\pm0.1$             &$(43-65)\%$           &$(95-97)\%$ \\

$J_{2,\mu}^{\Xi_b}$   &$\frac{3}{2}^-$          &$3.0$                &$4.1-4.5$            &$6.75\pm0.1$            &$(42-62)\%$            &$(93-96)\%$ \\  \hline

$\eta^{\Xi_b}$        &$\frac{1}{2}^-$         &$3.0$          &$4.0-4.4$            &$6.70\pm0.1$           &$(41-61)\%$            &$(90-93)\%$                            \\

$\eta_{1,\mu}^{\Xi_b}$   &$\frac{3}{2}^-$       &$3.0$            &$4.0-4.4$            &$6.70\pm0.1$              &$(41-61)\%$            &$(90-94)\%$ \\

$\eta_{2,\mu}^{\Xi_b}$   &$\frac{3}{2}^-$       &$3.0$              &$4.1-4.5$            &$6.75\pm0.1$              &$(41-62)\%$           &$(89-92)\%$ \\ \hline

\hline\hline
\end{tabular}
\caption{The energy scales $ \mu$, Borel windows $T^2$, continuum threshold parameters $s_0$,
 pole contributions and  perturbative contributions for the P-wave bottom baryon states.}\label{Borel}
\end{center}
\end{table}

\begin{table}
\begin{center}
\begin{tabular}{|c|c|c|c|c|c|c|c|c|}\hline\hline
\rm{Currents}     &$M (\rm{GeV})$   & $\lambda (10^{-1}\rm{GeV}^4) $&    $\rm{assignments} $
\\ \hline

$J^{\Lambda_b}$              &$5.91\pm0.13$       &$1.08\pm0.21$      &$\Lambda_b(5912)$        \\

$J_{1,\mu}^{\Lambda_b}$      &$5.91\pm0.14$       &$0.53\pm0.08$      &$\Lambda_b(5920)$      \\

$J_{2,\mu}^{\Lambda_b}$      &$5.92\pm0.15$       &$0.97\pm0.20$      &$\Lambda_b(5920)$   \\   \hline

$\eta^{\Lambda_b}$          &$5.91\pm0.13$      &$1.12\pm0.19$       &$\Lambda_b(5912)$        \\

$\eta_{1,\mu}^{\Lambda_b}$  &$5.91\pm0.13$      & $0.55\pm0.08$      &$\Lambda_b(5920)$                            \\

$\eta_{2,\mu}^{\Lambda_b}$  &$5.92\pm0.15$    & $0.96\pm0.20$       &$\Lambda_b(5920)$   \\   \hline

$J^{\Xi_b}$                 &$6.10\pm0.11$      & $1.59\pm0.25$      &$\Xi_b(6087)$                            \\

$J_{1,\mu}^{\Xi_b}$        &$6.10\pm0.10$        & $0.77\pm0.12$     &$\Xi_b(6095/6100)$     \\

$J_{2,\mu}^{\Xi_b}$       &$6.11\pm0.12$         & $1.43\pm0.22$       &$\Xi_b(6095/6100)$            \\  \hline

$\eta^{\Xi_b}$            &$6.09\pm0.11$        & $1.63\pm0.24$      &$\Xi_b(6087)$                            \\

$\eta_{1,\mu}^{\Xi_b}$    &$6.10\pm0.10$        & $0.79\pm0.11$      &$\Xi_b(6095/6100)$    \\

$\eta_{2,\mu}^{\Xi_b}$   &$6.12\pm0.13$         & $1.43\pm0.24$      &$\Xi_b(6095/6100)$ \\ \hline

\hline\hline
\end{tabular}
\caption{ The masses  and pole residues of the P-wave bottom baryon states with the possible assignments.} \label{mass}
\end{center}
\end{table}

\begin{table}
\begin{center}
\begin{tabular}{|c|c|c|c|c|c|c|c|c|}\hline\hline
\rm{Currents} &$J^P$ &$\mu$ & $T^2 (\rm{GeV}^2)$   &$\sqrt{s_0}(\rm GeV) $    &pole
&      $\rm{perturbative} $
\\ \hline

$J^{\Lambda_b}$             &$\frac{1}{2}^-$       &$2.9$            &$3.4-3.8$            &$6.65\pm0.1$               &$(52-74)\%$           &$(83-89)\%$       \\

$J_{1,\mu}^{\Lambda_b}$     &$\frac{3}{2}^-$       &$2.9$            &$3.4-3.8$         &$6.65\pm0.1$               &$(52-74)\%$           &$(83-89)\%$        \\

$J_{2,\mu}^{\Lambda_b}$     &$\frac{3}{2}^-$       &$2.9$            &$3.5-3.9$            &$6.70\pm0.1$               &$(54-75)\%$           &$(85-91)\%$    \\   \hline

$\eta^{\Lambda_b}$          &$\frac{1}{2}^-$       &$2.9$            &$3.4-3.8$            &$6.65\pm0.1$               &$(53-74)\%$           &$(80-85)\%$       \\

$\eta_{1,\mu}^{\Lambda_b}$  &$\frac{3}{2}^-$       &$2.9$            &$3.4-3.8$        &$6.65\pm0.1$               &$(53-75)\%$           &$(79-86)\%$    \\

$\eta_{2,\mu}^{\Lambda_b}$  &$\frac{3}{2}^-$       &$2.9$            &$3.5-3.9$           &$6.70\pm0.1$               &$(53-74)\%$           &$(79-86)\%$   \\   \hline

$J^{\Xi_b}$                 &$\frac{1}{2}^-$       &$3.0$            &$3.6-4.0$           &$6.80\pm0.1$               &$(55-75)\%$           &$(92-95)\%$    \\

$J_{1,\mu}^{\Xi_b}$         &$\frac{3}{2}^-$       &$3.0$            &$3.7-4.1$            &$6.80\pm0.1$               &$(53-73)\%$           &$(93-96)\%$  \\

$J_{2,\mu}^{\Xi_b}$         &$\frac{3}{2}^-$       &$3.0$            &$3.8-4.2$             &$6.85\pm0.1$               &$(54-74)\%$           &$(93-97)\%$   \\  \hline

$\eta^{\Xi_b}$              &$\frac{1}{2}^-$       &$3.0$            &$3.6-4.0$           &$6.80\pm0.1$               &$(56-76)\%$           &$(88-92)\%$    \\

$\eta_{1,\mu}^{\Xi_b}$      &$\frac{3}{2}^-$       &$3.0$            &$3.7-4.1$            &$6.80\pm0.1$               &$(54-74)\%$           &$(89-93)\%$    \\

$\eta_{2,\mu}^{\Xi_b}$      &$\frac{3}{2}^-$       &$3.0$            &$3.8-4.2$             &$6.85\pm0.1$               &$(53-73)\%$           &$(89-93)\%$     \\ \hline
\end{tabular}
\caption{The energy scales $\mu$, Borel windows $T^2$, continuum threshold parameters $s_0$,
 pole contributions ($> 50\%$)  and  perturbative contributions for the P-wave bottom baryon states.}\label{example-Borel}
\end{center}
\end{table}

\begin{table}
\begin{center}
\begin{tabular}{|c|c|c|c|c|c|c|c|c|}\hline\hline
\rm{Currents}     &$M (\rm{GeV})$   & $\lambda (10^{-1}\rm{GeV}^4) $&    $\rm{assignments}$     \\ \hline

$J^{\Lambda_b}$              &$5.92\pm0.15$        &$1.16\pm0.27$      &$\Lambda_b(5912)$        \\

$J_{1,\mu}^{\Lambda_b}$      &$5.92\pm0.13$        &$0.56\pm0.13$      &$\Lambda_b(5920)$                        \\

$J_{2,\mu}^{\Lambda_b}$      &$5.92\pm0.15$        &$1.01\pm0.25$      &$\Lambda_b(5920)$   \\   \hline

$\eta^{\Lambda_b}$           &$5.91\pm0.13$        &$1.16\pm0.25$      &$\Lambda_b(5912)$        \\

$\eta_{1,\mu}^{\Lambda_b}$   &$5.91\pm0.13$        &$0.57\pm0.11$      &$\Lambda_b(5920)$                            \\

$\eta_{2,\mu}^{\Lambda_b}$   &$5.93\pm0.16$        &$1.00\pm0.27$      &$\Lambda_b(5920)$   \\   \hline

$J^{\Xi_b}$                  &$6.10\pm0.12$        &$1.66\pm0.30$      &$\Xi_b(6087)$                            \\

$J_{1,\mu}^{\Xi_b}$          &$6.11\pm0.11$        &$0.83\pm0.13$      &$\Xi_b(6095/6100)$     \\

$J_{2,\mu}^{\Xi_b}$          &$6.11\pm0.12$        &$1.49\pm0.27$      &$\Xi_b(6095/6100)$            \\  \hline

$\eta^{\Xi_b}$               &$6.08\pm0.12$        &$1.65\pm0.25$      &$\Xi_b(6087)$                            \\

$\eta_{1,\mu}^{\Xi_b}$       &$6.10\pm0.11$        &$0.83\pm0.11$      &$\Xi_b(6095/6100)$    \\

$\eta_{2,\mu}^{\Xi_b}$       &$6.12\pm0.13$        &$1.48\pm0.30$      &$\Xi_b(6095/6100)$ \\
\hline\hline
\end{tabular}
\caption{The masses  and pole residues of the P-wave bottom baryon states (in the case of pole contributions $> 50\%$) with the possible assignments.}\label{example-mass}
\end{center}
\end{table}

\begin{figure}
 \centering
 \includegraphics[totalheight=5.5cm,width=7cm]{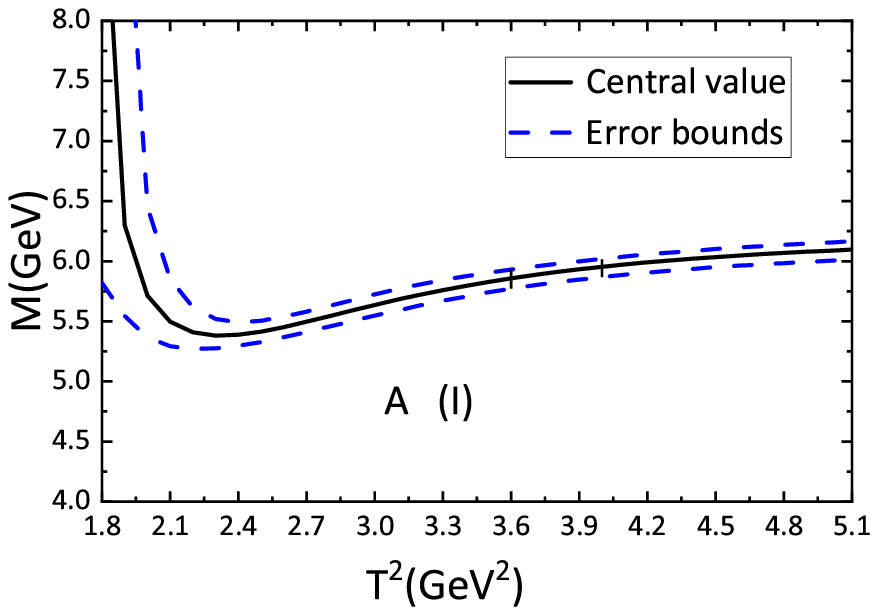}
 \includegraphics[totalheight=5.5cm,width=7cm]{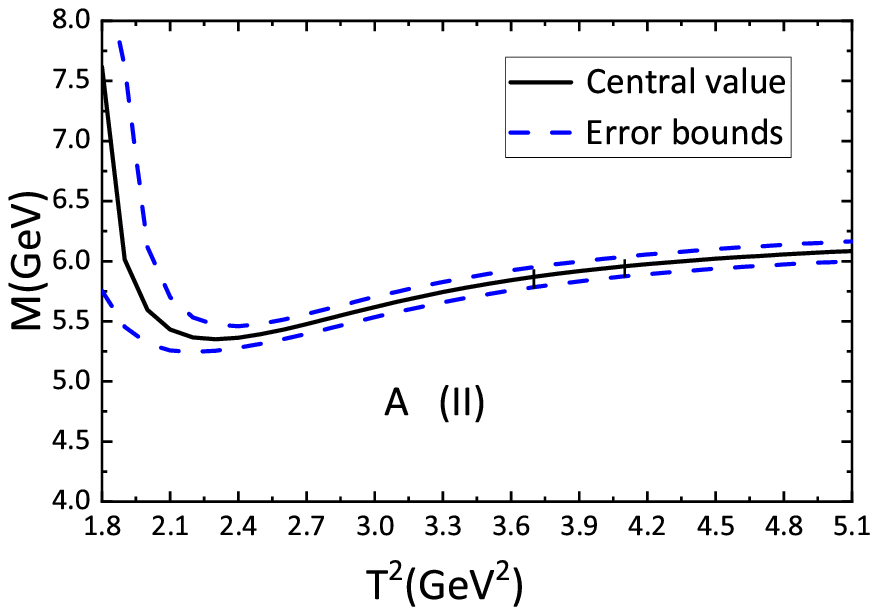}
  \includegraphics[totalheight=5.5cm,width=7cm]{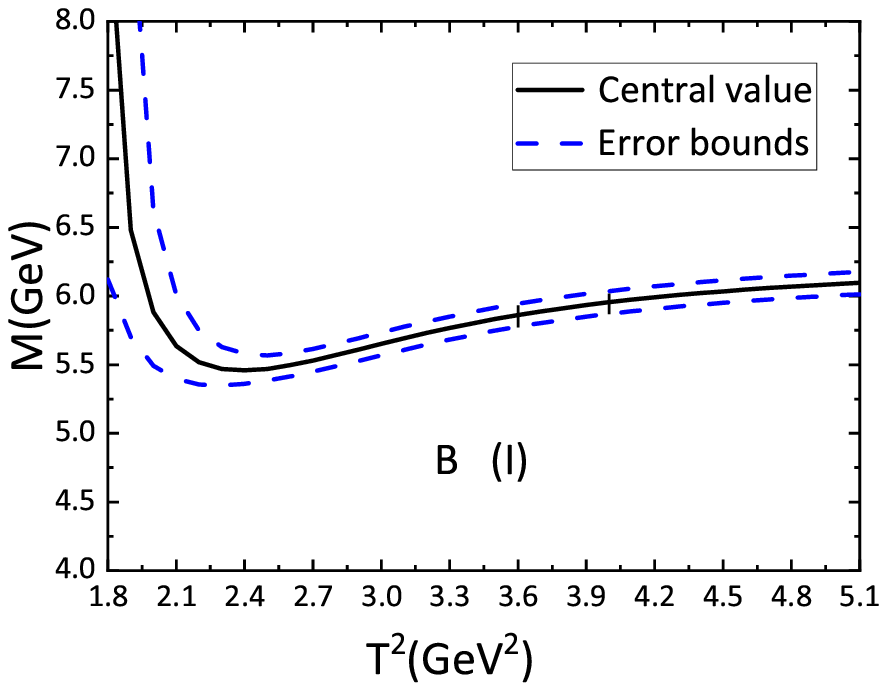}
  \includegraphics[totalheight=5.5cm,width=7cm]{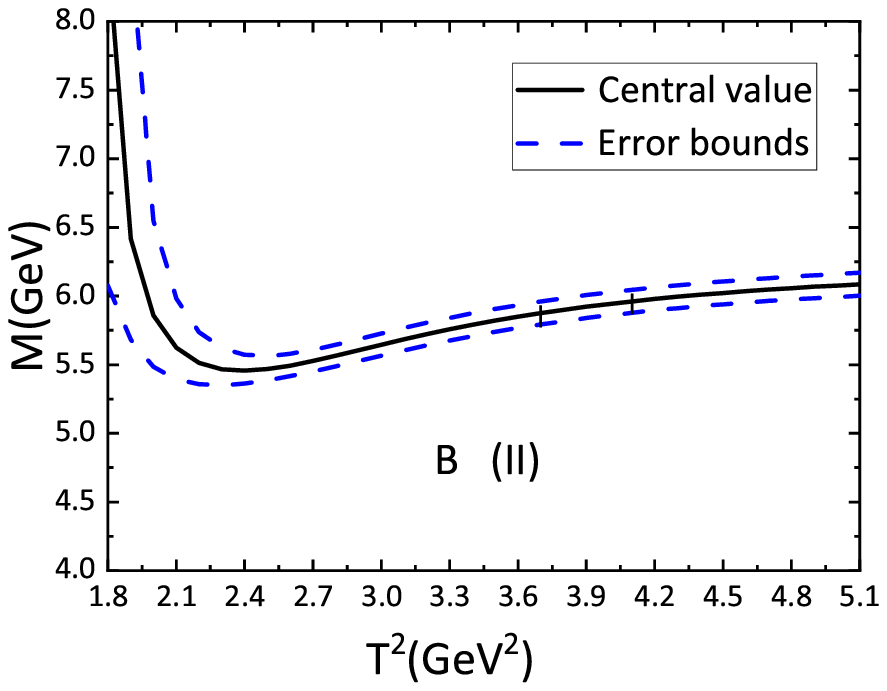}
  \includegraphics[totalheight=5.5cm,width=7cm]{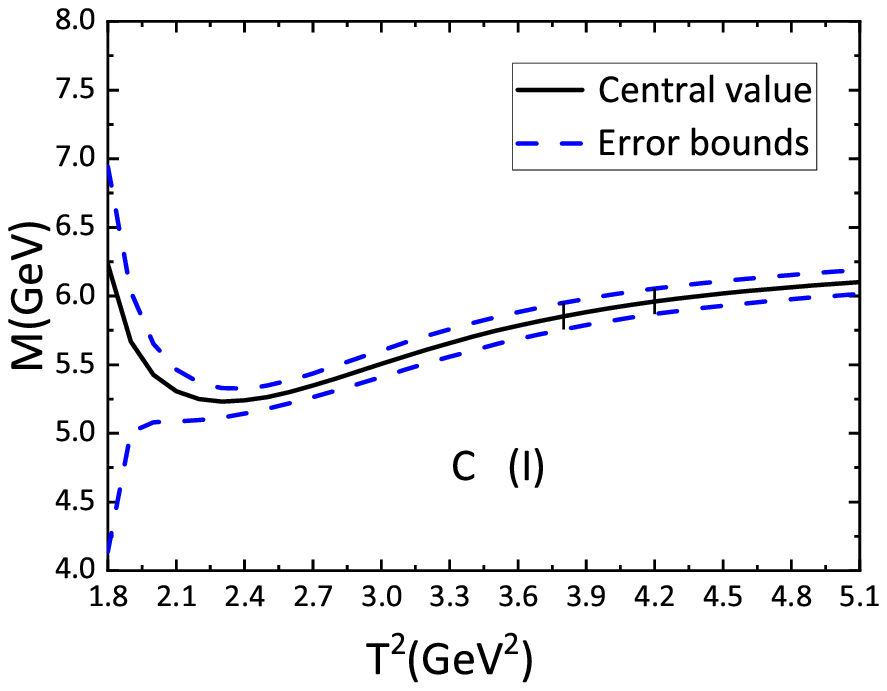}
  \includegraphics[totalheight=5.5cm,width=7cm]{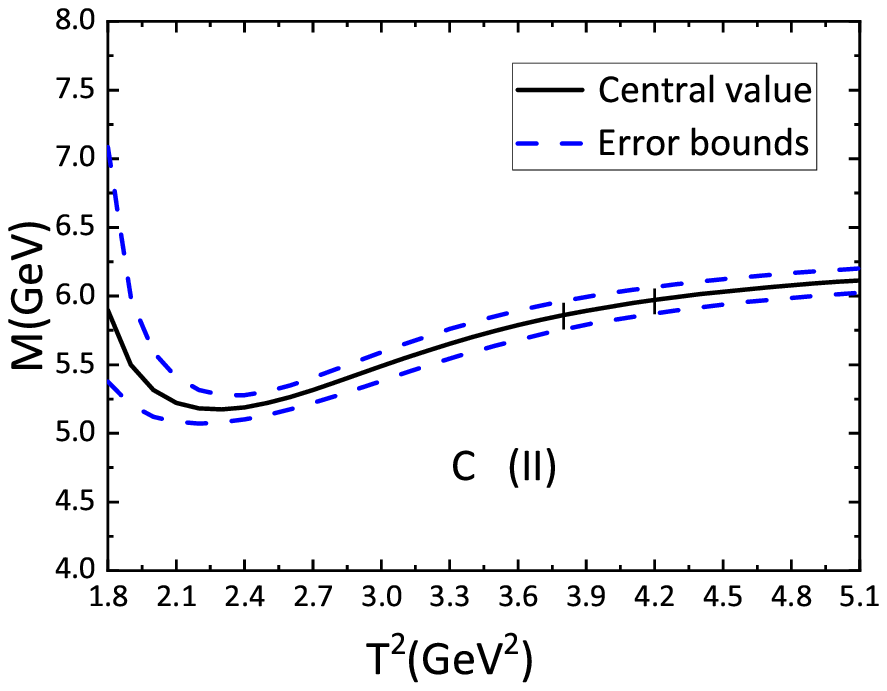}
 \caption{ The masses  of the  ${\Lambda_b}({\frac{1}{2}^-})$, ${\Lambda_b}(\frac{3}{2}^-,1)$ and ${\Lambda_b}(\frac{3}{2}^-,2)$ states (labeled as $A$, $B$ and $C$ respectively) with variations of the Borel parameters $T^2$, where the (I) and (II)  denote  the currents  with the partial and covariant derivatives,  respectively.}\label{mass-lambda}
\end{figure}

\begin{figure}
 \centering
  \includegraphics[totalheight=5.5cm,width=7cm]{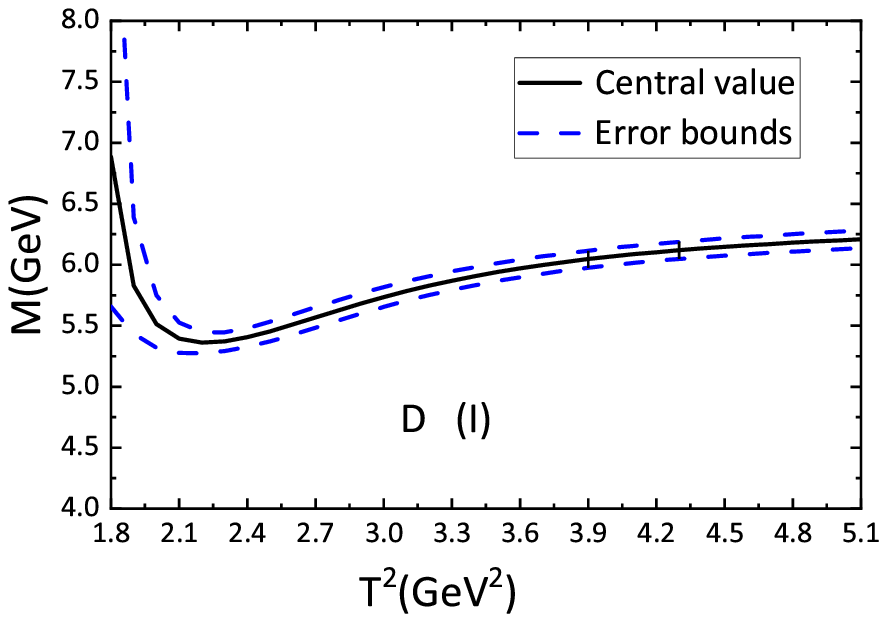}
  \includegraphics[totalheight=5.5cm,width=7cm]{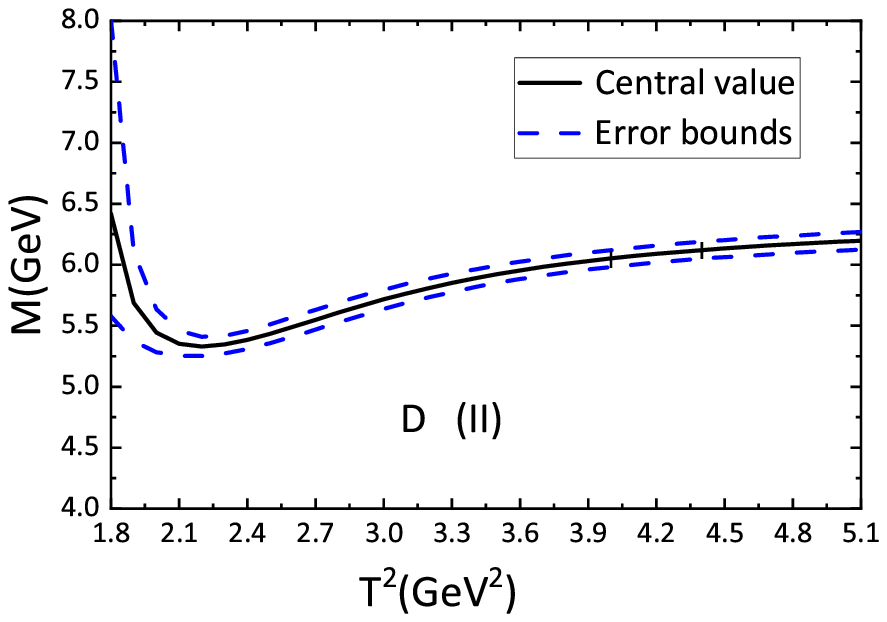}
  \includegraphics[totalheight=5.5cm,width=7cm]{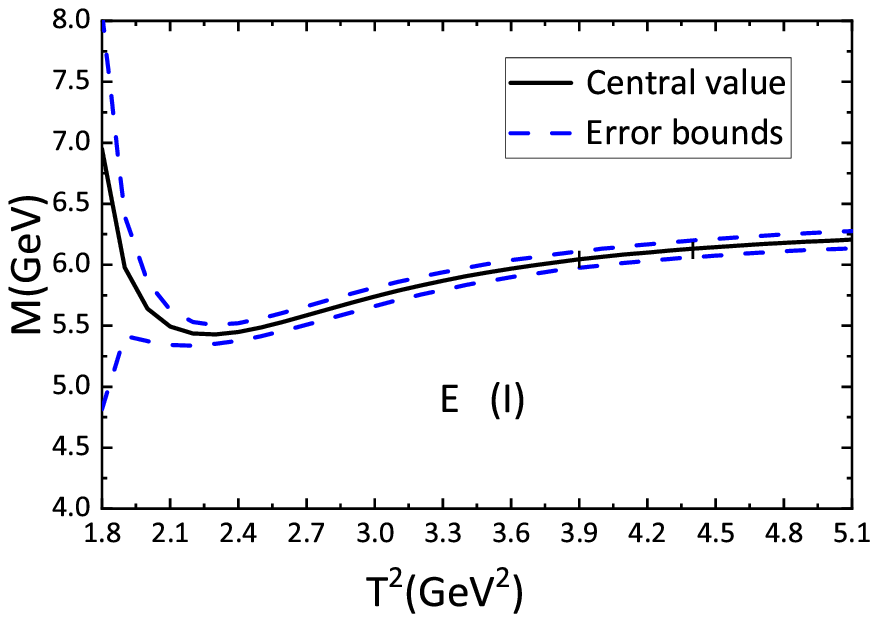}
  \includegraphics[totalheight=5.5cm,width=7cm]{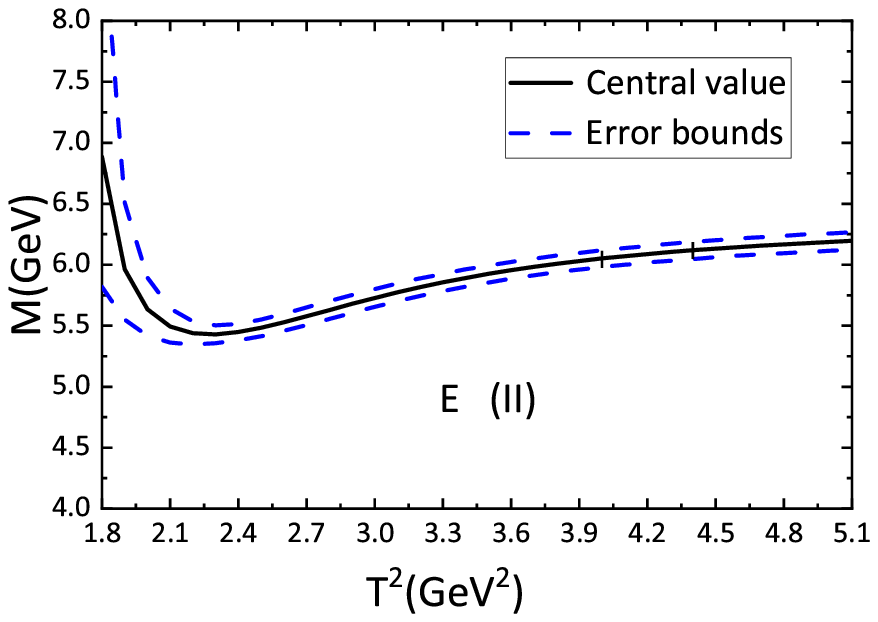}
  \includegraphics[totalheight=5.5cm,width=7cm]{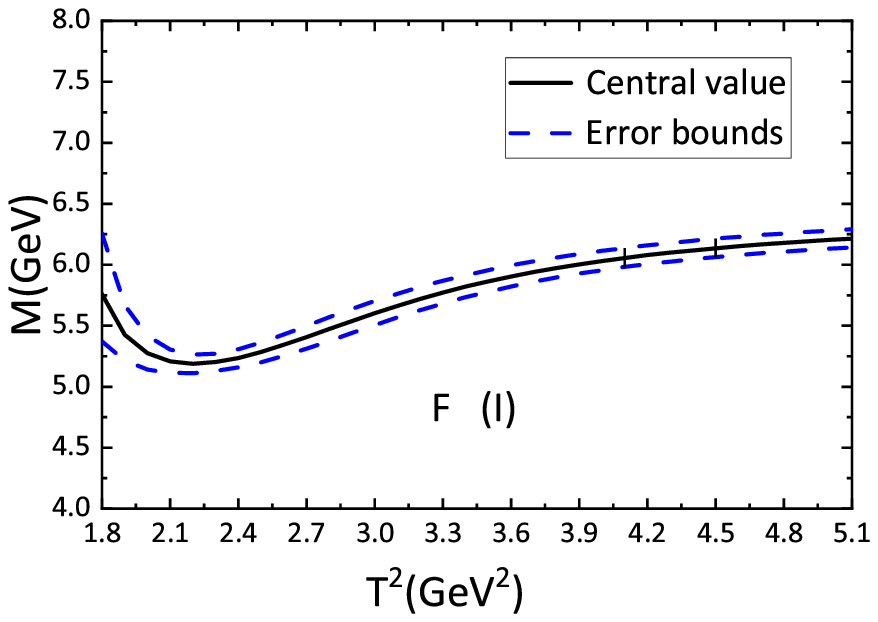}
  \includegraphics[totalheight=5.5cm,width=7cm]{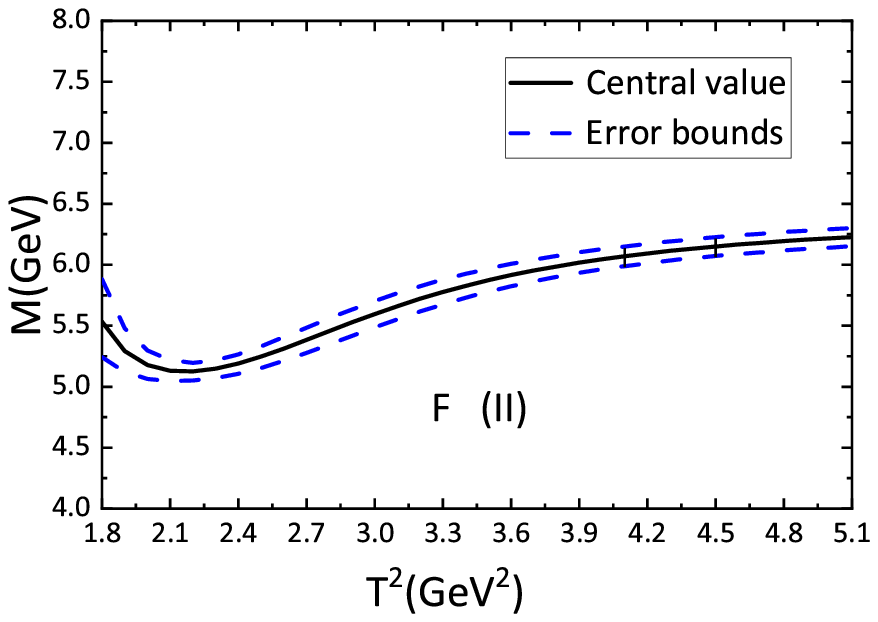}
 \caption{ The masses  of the  ${\Xi_b}({\frac{1}{2}^-})$, ${\Xi_b}(\frac{3}{2}^-,1)$ and ${\Xi_b}(\frac{3}{2}^-,2)$ states (labeled as $D$, $E$ and $F$ respectively) with variations of the Borel parameters $T^2$, where the (I) and (II)  denote the currents  with the partial and covariant derivatives, respectively.}\label{mass-Xi}
\end{figure}

In Figs.\ref{mass-lambda}-\ref{mass-Xi}, we plot the variation trends of the baryon masses  with the Borel parameters, where  the two vertical lines indicate the ranges of the Borel platforms. In Table \ref{mass}, we show the baryon masses and pole residues explicitly via accounting all uncertainties of the input parameters. As can be seen, both the currents containing the covariant derivatives and partial derivatives can lead to the baryon masses consistent with the experimental data.

The predicted masses from the currents with partial (covariant) derivatives  $M_{-}=5.91  {\pm0.13}\,\rm{GeV}$ ($5.91 {\pm0.13}\,\rm{GeV}$), $5.91 {\pm0.14}\,\rm{GeV}$ ($5.91 {\pm0.13}\,\rm{GeV}$) and $5.92 {\pm0.15}\,\rm{GeV}$ ($5.92 {\pm0.15}\,\rm{GeV}$) are consistent with the experimentally measured masses $M_{\Lambda_b(5912)}=5911.97\pm0.12 \pm0.02 \pm0.66 \mbox{ MeV}$ or $M_{\Lambda_b(5920)}=5919.77 \pm0.08 \pm0.02 \pm0.66 \mbox{ MeV}$ from the LHCb collaboration  \cite{LHCb-5912-5920}. The numerical results indicate that the $\Lambda_b(5912)$ and $\Lambda_b(5920)$ could be the $\Lambda$-type  P-wave  bottom baryon states  with the $J^P=\frac{1}{2}^-$ and $\frac{3}{2}^-$, respectively. Analogously, the predicted masses $M=6.10  {\pm0.11}\,\rm{GeV}$ ($6.09 {\pm0.11}\,\rm{GeV}$), $6.10 {\pm0.10}\,\rm{GeV}$ ($6.10 {\pm0.10}\,\rm{GeV}$) and $6.11 {\pm0.12}\,\rm{GeV}$ ($6.12 {\pm0.13}\,\rm{GeV}$) are consistent with the experimentally measured masses  $M_{\Xi_b(6087)}=6087.24\pm0.2 \pm0.06 \pm0.5 \mbox{ MeV}$,  $M_{\Xi_b(6095)}=6095.26 \pm0.15 \pm0.03 \pm0.5 \mbox{ MeV}$, or $M_{\Xi_b(6100)}=6099.74\pm0.11\pm0.02\pm0.6 \mbox{ MeV}$ from the LHCb collaboration \cite{LHCb-6087-6095-GKA}.  Our numerical results indicate that the $\Xi_b(6087)$  and $\Xi_b(6095/6100)$ could be the  $\Lambda$-type P-wave bottom baryon states with the $J^P=\frac{1}{2}^-$ and $\frac{3}{2}^-$, respectively. The $\Lambda_b(5920)$ and $\Xi_b(6095/6100)$ could be interpreted at least have two remarkable under-structures or Fock components.

We cannot assign those bottom baryon states unambiguously  with the masses alone, at least, the dominant strong decays should be investigated, such as
\begin{eqnarray}
\Xi_b^0(6087)  &\to& \Xi_b^{\prime-}\pi^+/\Xi_b^{0}\rho^0  \to \Xi_b^0 \pi^-\pi^+\, , \nonumber\\
\Xi_b^-(?)&\to& \Xi_b^{\prime0}\pi^-/\Xi_b^{-}\rho^0 \to \Xi_b^-\pi^+\pi^-\, , \nonumber \\
\Xi_b^0(6095)&\to& \Xi_b^{*-}\pi^+/\Xi_b^{0}\rho^0 \to \Xi_b^0\pi^-\pi^+\, , \nonumber \\
\Xi_b^-(6100)&\to& \Xi_b^{*0}\pi^- /\Xi_b^{-}\rho^0\to \Xi_b^-\pi^+\pi^-\, ,  
\end{eqnarray}
\begin{eqnarray}
\Lambda_b^0(5912)  &\to& \Sigma_b^{-}\pi^+/\Lambda_b^{0}\rho^0   \to \Lambda_b^0 \pi^-\pi^+\, , \nonumber \\
\Lambda_b^0(5920)  &\to& \Sigma_b^{*-}\pi^+/\Lambda_b^{0}\rho^0   \to \Lambda_b^0 \pi^-\pi^+\, ,
\end{eqnarray}
where the intermediate $\rho^0$ is off-shell, and the $\Xi_b^{\prime0}$ is also off-shell as the decay $\Xi_b^{\prime0}\to \Xi_b^-\pi^+$ is kinematically forbidden.
At the present time, there only exist experimental evidences for  the isospin doublet $(\Xi_b^0(6095),\Xi_b^-(6100))$, while there still lack experimental evidences for the isospin doublet $(\Xi_b^0(6087), \Xi_b^-(?))$.
We can explore those three-body decays with the (light-cone) QCD sum rules directly or indirectly \cite{Aliev-Sigmab,CHX-6100,WZG-NPB-4500}, then confront the predictions to the experimental data in the futures to diagnose the nature of those P-wave baryon states, it will  be our next work.

\section{Conclusion}
In this work, we extend our previous works to explore the $\Lambda$-type P-wave bottom baryon states with the QCD sum rules in details. We introduce a relative P-wave between the two light quarks in the diquarks to construct the interpolating currents, and refer to them as the $\Lambda$-type currents because the two light quarks are antisymmetric. We carry out the operator product expansion up to the vacuum condensates of dimension 10 in a self-consistent way,   obtain the spectral representations through dispersion relation, and distinguish the contributions of the negative-parity and positive-parity bottom baryon states unambiguously,  then
determine the ideal  energy scales of the QCD spectral densities using the modified energy scale formula by considering the light-flavor $SU(3)$ breaking effects. Our numerical results support assigning  the $\Lambda^0_b(5912)$ and $\Xi_b^0(6087)$   to be the $\Lambda$-type P-wave   baryon states with the spin-parity $J^P=\frac{1}{2}^-$ and valence quarks $udb$ and $usb$, respectively, and assigning the $\Lambda^0_b(5920)$ and $\Xi^0_b(6095)$ ($\Xi^-_b(6100)$)  to be the $\Lambda$-type P-wave   baryon states with the spin-parity $J^P=\frac{3}{2}^-$ and valence quarks $udb$ and $usb$ ($dsb$), respectively. The $\Xi^0_b(6095)$ and $\Xi^-_b(6100)$ form an isospin doublet, while the isospin partner of the $\Xi_b^0(6087)$ has not been observed yet.
The $\Xi_b(6095)$ and $\Xi_b(6100)$ maybe have two structures or Fock components, as there exist  two $J^P=\frac{3}{2}^-$ currents with different structures but couple potentially to the bottom baryon states with almost degenerated masses. Furthermore, we observe that the currents with the covariant or partial derivatives lead to almost the same baryon masses, if only the baryon masses are concerned,  we can choose either the covariant or partial derivatives in constructing the currents. According to the quantum field theory, we construct gauge invariant  currents with the same quantum numbers as the hadrons to interpolate them, therefore  the covariant  derivatives are preferred.

\section*{Appendix}
The QCD spectral densities $\rho_{j,QCD}^0(s)$ and $\rho_{j,QCD}^1(s)$ for the currents with the partial derivatives,
\begin{eqnarray}
\rho_{j,QCD}^0(s)&=&\rho_{\frac{1}{2},\Lambda_b}^0(s)\, ,\,\, \rho_{\frac{3}{2},1,\Lambda_b}^0(s)\, , \, \,\rho_{\frac{3}{2},2,\Lambda_b}^0(s)\, , \, \,\rho_{\frac{1}{2},\Xi_b}^0(s)\, ,\,\, \rho_{\frac{3}{2},1,\Xi_b}^0(s)\, , \, \,\rho_{\frac{3}{2},2,\Xi_b}^0(s)\, , \nonumber \\
\rho_{j,QCD}^1(s)&=&\rho_{\frac{1}{2},\Lambda_b}^1(s)\, ,\,\, \rho_{\frac{3}{2},1,\Lambda_b}^1(s)\, , \, \,\rho_{\frac{3}{2},2,\Lambda_b}^1(s)\, , \, \,\rho_{\frac{1}{2},\Xi_b}^1(s)\, ,\,\, \rho_{\frac{3}{2},1,\Xi_b}^1(s)\, , \, \,\rho_{\frac{3}{2},2,\Xi_b}^1(s)\, ,
\end{eqnarray}
where $j=\frac{1}{2}$, $\frac{3}{2}$,

\begin{eqnarray}
\rho_{\frac{1}{2},\Lambda_b}^0(s)&=&\rho_{\frac{1}{2},\Xi_b}^0(s)\mid_{m_s \to 0, \langle\bar{s}s\rangle \to\langle\bar{q}q\rangle, \langle\bar{s}g_s\sigma Gs\rangle \to\langle\bar{q}g_s\sigma Gq\rangle}\nonumber \, , \\
\rho_{\frac{3}{2},1,\Lambda_b}^0(s)&=&\rho_{\frac{3}{2},1,\Xi_b}^0(s)\mid_{m_s \to 0, \langle\bar{s}s\rangle \to\langle\bar{q}q\rangle, \langle\bar{s}g_s\sigma Gs\rangle \to\langle\bar{q}g_s\sigma Gq\rangle}\nonumber \, , \\
\rho_{\frac{3}{2},2,\Lambda_b}^0(s)&=&\rho_{\frac{3}{2},2,\Xi_b}^0(s)\mid_{m_s \to 0, \langle\bar{s}s\rangle \to\langle\bar{q}q\rangle, \langle\bar{s}g_s\sigma Gs\rangle \to\langle\bar{q}g_s\sigma Gq\rangle}\, ,
\end{eqnarray}

\begin{eqnarray}
\rho_{\frac{1}{2},\Lambda_b}^1(s)&=&\rho_{\frac{1}{2},\Xi_b}^1(s)\mid_{m_s \to 0, \langle\bar{s}s\rangle \to\langle\bar{q}q\rangle, \langle\bar{s}g_s\sigma Gs\rangle \to\langle\bar{q}g_s\sigma Gq\rangle}\nonumber \, , \\
\rho_{\frac{3}{2},1,\Lambda_b}^1(s)&=&\rho_{\frac{3}{2},1,\Xi_b}^1(s)\mid_{m_s \to 0, \langle\bar{s}s\rangle \to\langle\bar{q}q\rangle, \langle\bar{s}g_s\sigma Gs\rangle \to\langle\bar{q}g_s\sigma Gq\rangle}\nonumber \, , \\
\rho_{\frac{3}{2},2,\Lambda_b}^1(s)&=&\rho_{\frac{3}{2},2,\Xi_b}^1(s)\mid_{m_s \to 0, \langle\bar{s}s\rangle \to\langle\bar{q}q\rangle, \langle\bar{s}g_s\sigma Gs\rangle \to\langle\bar{q}g_s\sigma Gq\rangle}\, ,
\end{eqnarray}

\begin{eqnarray}
\rho^0_{\frac{1}{2},\Xi_b}(s)&=&\frac{m_b} {32\pi^4} \int_{x_i}^{1}dx(1-x)^3({s-\tilde{m}_b^2})^3-\frac{m_b^3} {96\pi^2} \langle\frac{\alpha_{s}GG}{\pi}\rangle\int_{x_i}^{1}dx\frac{(1-x)^3}{x^3}\nonumber\\
&&+\frac{m_b} {32\pi^2} \langle\frac{\alpha_{s}GG}{\pi}\rangle\int_{x_i}^{1}dx\frac{(x-1)(x+1)(3x-2)}{x^2}(s-\tilde{m}_b^2)\nonumber\\
&&+m_sm_b\langle\bar{s}s\rangle\langle\frac{\alpha_{s}GG}{\pi}\rangle\int_{x_i}^{1}dx\frac{1}{32x}\delta(s-\tilde{m}_b^2)\nonumber\\
&&+\frac{m_sm_b(5\langle\bar{s}g_s\sigma Gs\rangle-12\langle\bar{q}g_s\sigma Gq\rangle)}{32\pi^2}\int_{x_i}^{1}dx\nonumber\\
&& +\frac{m_b(\langle\bar{q}q\rangle\langle\bar{s}g_s\sigma Gs\rangle+\langle\bar{s}s\rangle\langle\bar{q}g_s\sigma Gq\rangle)}{4}\delta(s-m_b^2)\nonumber\\
&&-\frac{m_b\langle\bar{s}g_s\sigma Gs\rangle\langle\bar{q}g_s\sigma Gq\rangle}{48T^2}\left(-2+\frac{3s}{T^2}\right)\delta(s-m_b^2)  \, ,
\end{eqnarray}

\begin{eqnarray}
\rho^1_{\frac{1}{2},\Xi_b}(s)&=&\frac{1} {16\pi^4} \int_{x_i}^{1}dxx(1-x)^3({s-\tilde{m}_b^2})^3+\frac{3m_s(2\langle\bar{s}s\rangle-\langle\bar{q}q\rangle)} {4\pi^2}\int_{x_i}^{1}dxx(1-x)(s-\tilde{m}_b^2)\nonumber\\
&&-\frac{m_b^2} {48\pi^2} \langle\frac{\alpha_{s}GG}{\pi}\rangle\int_{x_i}^{1}dx\frac{(1-x)^3}{x^2}+\frac{3}{64\pi^2}\langle\frac{\alpha_{s}GG}{\pi}\rangle\int_{x_i}^{1}dx(1-x)^2(s-\tilde{m}_b^2)
\nonumber\\
&&+\frac{m_sm_b^2(\langle\bar{q}q\rangle-2\langle\bar{s}s\rangle)}{24 T^2}\langle\frac{\alpha_{s}GG}{\pi}\rangle\int_{x_i}^{1}dx\frac{(1-x)}{x^2}\delta(s-\tilde{m}_b^2)\nonumber\\
&&+\frac{m_s\langle\bar{s}s\rangle}{32}\langle\frac{\alpha_{s}GG}{\pi}\rangle\int_{x_i}^{1}dx\delta(s-\tilde{m}_b^2)+\frac{m_s\langle\bar{q}q\rangle} {48} \langle\frac{\alpha_{s}GG}{\pi}\rangle\delta(s-m_b^2)\nonumber\\
&&+\frac{m_s\langle\bar{q}g_s\sigma Gq\rangle} {16\pi^2}\int_{x_i}^{1}dx(7x-1)-\frac{11m_s\langle\bar{s}g_s\sigma Gs\rangle}{32\pi^2}\int_{x_i}^{1}dxx\nonumber\\
&&-\frac{\langle\bar{s}g_s\sigma Gs\rangle\langle\bar{q}g_s\sigma Gq\rangle}{16T^2}\left(1+\frac{s}{T^2}\right)\delta(s-m_b^2) \, ,
\end{eqnarray}

\begin{eqnarray}
\rho^0_{\frac{3}{2},1,\Xi_b}(s)&=&\frac{m_b} {128\pi^4} \int_{x_i}^{1}dx(1-x)^3({s-\tilde{m}_b^2})^3-\frac{m_b^3} {384\pi^2} \langle\frac{\alpha_{s}GG}{\pi}\rangle\int_{x_i}^{1}dx\frac{(1-x)^3}{x^3}\nonumber\\
&&-\frac{m_b}{576\pi^2} \langle\frac{\alpha_{s}GG}{\pi}\rangle\int_{x_i}^{1}dx\frac{(1-x)^3}{x}(-3s+2\tilde{m}_b^2)\nonumber\\
&&-\frac{m_b} {768\pi^2} \langle\frac{\alpha_{s}GG}{\pi}\rangle\int_{x_i}^{1}dx\frac{(x-1)(6-13x+x^2)}{x^2}(s-\tilde{m}_b^2)\nonumber\\
&&-\frac{m_sm_b\langle\bar{s}s\rangle}{384}\langle\frac{\alpha_{s}GG}{\pi}\rangle\int_{x_i}^{1}dx\left[\frac{3-4x}{x}-\frac{4(x-1)}{x}\frac{s}{T^2}\right]\delta(s-\tilde{m}_b^2)\nonumber\\
&&+\frac{5m_sm_b\langle\bar{s}g_s\sigma Gs\rangle}{128\pi^2}\int_{x_i}^{1}dx+\frac{m_sm_b\langle\bar{q}g_s\sigma Gq\rangle}{192\pi^2}\int_{x_i}^{1}dx(x-19)\nonumber\\
&&+\frac{m_sm_b\langle\bar{q}g_s\sigma Gq\rangle}{96\pi^2}\int_{x_i}^{1}dx\frac{s(1-x)^2}{x}\delta(s-\tilde{m}_b^2)\nonumber\\
&&+\frac{m_b(\langle\bar{q}q\rangle\langle\bar{s}g_s\sigma Gs\rangle+\langle\bar{s}s\rangle\langle\bar{q}g_s\sigma Gq\rangle)}{16}\delta(s-m_b^2)\nonumber\\
&&-\frac{m_b\langle\bar{s}g_s\sigma Gs\rangle\langle\bar{q}g_s\sigma Gq\rangle}{256}\delta(s-m_b^2)
 \, ,
\end{eqnarray}

\begin{eqnarray}
\rho^1_{\frac{3}{2},1,\Xi_b}(s)&=&\frac{1} {64\pi^4} \int_{x_i}^{1}dxx(1-x)^3({s-\tilde{m}_b^2})^3\nonumber\\
&&-\frac{3m_s(2\langle\bar{q}q\rangle-2\langle\bar{s}s\rangle)} {16\pi^2}\int_{x_i}^{1}dxx(1-x)(s-\tilde{m}_b^2)\nonumber\\
&&-\frac{m_b^2} {192\pi^2} \langle\frac{\alpha_{s}GG}{\pi}\rangle\int_{x_i}^{1}dx\frac{(1-x)^3}{x^2}-\frac{1}{256\pi^2}\langle\frac{\alpha_{s}GG}{\pi}\rangle\int_{x_i}^{1}dx(1-x)^2(s-\tilde{m}_b^2)\nonumber\\
&&+\frac{m_sm_b^2(\langle\bar{q}q\rangle-2\langle\bar{s}s\rangle)}{96 T^2}\langle\frac{\alpha_{s}GG}{\pi}\rangle\int_{x_i}^{1}dx(1-x)\delta(s-\tilde{m}_b^2)\nonumber\\
&&-\frac{m_s\langle\bar{s}s\rangle}{384}\langle\frac{\alpha_{s}GG}{\pi}\rangle\int_{x_i}^{1}dx\delta(s-\tilde{m}_b^2)
+\frac{m_s\langle\bar{q}q\rangle} {192} \langle\frac{\alpha_{s}GG}{\pi}\rangle\delta(s-m_b^2)
\nonumber\\
&&+\frac{m_s\langle\bar{q}g_s\sigma Gq\rangle} {384\pi^2}\int_{x_i}^{1}dx(2+x)-\frac{11m_s\langle\bar{s}g_s\sigma Gs\rangle}{128\pi^2}\int_{x_i}^{1}dxx\nonumber\\
&&-\frac{3\langle\bar{s}g_s\sigma Gs\rangle\langle\bar{q}g_s\sigma Gq\rangle}{64T^2}\left(1-\frac{s}{T^2}\right)\delta(s-m_b^2) \, ,
\end{eqnarray}

\begin{eqnarray}
\rho^0_{\frac{3}{2},2,\Xi_b}(s)&=&\frac{m_b} {192\pi^4} \int_{x_i}^{1}dx(4+x)(1-x)^3(s-\tilde{m}_b^2)^3\nonumber\\
&&+\frac{m_sm_b(\langle\bar{s}s\rangle-2\langle\bar{q}q\rangle)}{8\pi^2}\int_{x_i}^{1}dxx(1-x)(s-\tilde{m}_b^2)\nonumber\\
&&+\frac{m_b^3}{576\pi^2}\langle\frac{\alpha_{s}GG}{\pi}\rangle\int_{x_i}^{1}dx\frac{(x-1)^3(4+x)} {x^3}\nonumber\\
&&-\frac{m_b}{192\pi^2}\langle\frac{\alpha_{s}GG}{\pi}\rangle\int_{x_i}^{1}dx\frac{(x-1)^3(4+x)} {x^2}(s-\tilde{m}_b^2)\nonumber\\
&&+\frac{m_b}{384\pi^2}\langle\frac{\alpha_{s}GG}{\pi}\rangle\int_{x_i}^{1}dx\frac{(x-1)[s(24-51x+44x^2)+(-10+23x-30x^2)\tilde{m}_b^2]}{x}\nonumber\\
&&+\frac{m_sm_b(\langle\bar{s}s\rangle-2\langle\bar{q}q\rangle)}{48}\langle\frac{\alpha_{s}GG}{\pi}\rangle\int_{x_i}^{1}dx\frac{(1-x)(3x-1)}{3x^2}\delta(s-\tilde{m}_b^2)\nonumber\\
&&-\frac{m_sm_b\langle\bar{q}q\rangle}{72}\langle\frac{\alpha_{s}GG}{\pi}\rangle\int_{x_i}^{1}dx\delta(s-\tilde{m}_b^2)+\frac{m_sm_b\langle\bar{q}q\rangle}{144}\langle\frac{\alpha_{s}GG}{\pi}\rangle\delta(s-m_b^2)\nonumber\\
&&+\frac{m_sm_b\langle\bar{s}s\rangle}{288}\langle\frac{\alpha_{s}GG}{\pi}\rangle\int_{x_i}^{1}dx\left(1+\frac{x-1}{x}\frac{s}{T^2}\right)\delta(s-\tilde{m}_b^2)\nonumber\\
&&+\frac{7m_sm_b\langle\bar{q}g_s\sigma Gq\rangle}{96\pi^2}\int_{x_i}^{1}dx\frac{s(1-x)^2}{x}\delta(s-\tilde{m}_b^2)\nonumber\\
&&+\frac{m_sm_b\langle\bar{s}g_s\sigma Gs\rangle}{384\pi^2}\int_{x_i}^{1}dx(81-64x)+\frac{m_sm_b\langle\bar{q}g_s\sigma Gq\rangle}{384\pi^2}\int_{x_i}^{1}dx(-842+782x)\nonumber\\
&&+\frac{3m_b(\langle\bar{q}q\rangle\langle\bar{s}g_s\sigma Gs\rangle+\langle\bar{s}s\rangle\langle\bar{q}g_s\sigma Gq\rangle)}{16}\delta(s-m_b^2)\nonumber\\
&&-\frac{m_b\langle\bar{s}g_s\sigma Gs\rangle\langle\bar{q}g_s\sigma Gq\rangle}{96T^2}\left(-7+\frac{13s}{2T^2}\right)\delta(s-m_b^2) \, ,
\end{eqnarray}

\begin{eqnarray}
\rho^1_{\frac{3}{2},2,\Xi_b}(s)&=&\frac{1} {64\pi^4} \int_{x_i}^{1}dxx(2+x)(1-x)^3(s-\tilde{m}_b^2)^3+\frac{m_b^2} {192\pi^2} \langle\frac{\alpha_{s}GG}{\pi}\rangle\int_{x_i}^{1}dx\frac{(x-1)^3(2+x)}{x^2}\nonumber\\
&&+\frac{m_s\langle\bar{q}q\rangle}{8\pi^2}\int_{x_i}^{1}dxx(x-1)(x+3)(s-\tilde{m}_b^2)+\frac{m_s\langle\bar{q}q\rangle}{72}\langle\frac{\alpha_{s}GG}{\pi}\rangle\delta(s-m_b^2)\nonumber\\
&&-\frac{m_s\langle\bar{s}s\rangle}{8\pi^2}\int_{x_i}^{1}dxx(x-1)(8x-1)(s-\tilde{m}_b^2)\nonumber\\
&&+\langle\frac{\alpha_{s}GG}{\pi}\rangle\int_{x_i}^{1}dx\frac{(1-x)(20-34x-13x^2)}{1152\pi^2}(s-\tilde{m}_b^2)\nonumber\\
&&-\frac{m_sm_b^2\langle\bar{s}s\rangle}{144T^2}\langle\frac{\alpha_{s}GG}{\pi}\rangle\int_{x_i}^{1}dx\frac{(1-x)(8x-1)}{x^2}\delta(s-\tilde{m}_b^2)\nonumber\\
&&-\frac{m_sm_b^2\langle\bar{q}q\rangle}{144T^2}\langle\frac{\alpha_{s}GG}{\pi}\rangle\int_{x_i}^{1}dx\frac{(x-1)(x+3)}{x^2}\delta(s-\tilde{m}_b^2)\nonumber\\
&&-\frac{m_s\langle\bar{q}q\rangle}{144}\langle\frac{\alpha_{s}GG}{\pi}\rangle\int_{x_i}^{1}dxx\delta(s-\tilde{m}_b^2)+\frac{5m_s\langle\bar{s}s\rangle}{576}\langle\frac{\alpha_{s}GG}{\pi}\rangle\int_{x_i}^{1}dxx\delta(s-\tilde{m}_b^2)\nonumber\\
&&-\frac{m_s\langle\bar{s}g_s\sigma Gs\rangle}{128\pi^2}\int_{x_i}^{1}dxx(68x-47)+\frac{m_s\langle\bar{q}g_s\sigma Gq\rangle}{192\pi^2}\int_{x_i}^{1}dxx(29x-11)\nonumber\\
&&+\frac{5(\langle\bar{q}q\rangle\langle\bar{s}g_s\sigma Gs\rangle+\langle\bar{s}s\rangle\langle\bar{q}g_s\sigma Gq\rangle)}{48}\delta(s-m_b^2)\nonumber\\
&&+\frac{\langle\bar{s}g_s\sigma Gs\rangle\langle\bar{q}g_s\sigma Gq\rangle}{576T^2}\left(37-\frac{13s}{3T^2}\right)\delta(s-m_b^2) \, ,
\end{eqnarray}
where $\tilde{m}_b^2=\frac{{m}_b^2}{x}$, $x_i=\frac{{m}_b^2}{s}$.

With the simple replacements,
\begin{eqnarray}
\rho_{j,QCD}^0(s)&\to&\rho_{j,QCD}^0(s)+\tilde{\rho}^0_{j,QCD}(s)\, , \nonumber \\
\rho_{j,QCD}^1(s)&\to&\rho_{j,QCD}^1(s)+\tilde{\rho}^1_{j,QCD}(s)\, ,
\end{eqnarray}
we obtain the corresponding QCD spectral densities for the currents with the covariant derivatives, where the additional terms,

\begin{eqnarray}
\tilde{\rho}_{j,QCD}^0(s)&=&\tilde{\rho}_{\frac{1}{2},\Lambda_b}^0(s)\, ,\,\, \tilde{\rho}_{\frac{3}{2},1,\Lambda_b}^0(s)\, , \, \,\tilde{\rho}_{\frac{3}{2},2,\Lambda_b}^0(s)\, , \, \,\tilde{\rho}_{\frac{1}{2},\Xi_b}^0(s)\, ,\,\, \tilde{\rho}_{\frac{3}{2},1,\Xi_b}^0(s)\, , \, \,\tilde{\rho}_{\frac{3}{2},2,\Xi_b}^0(s)\, , \nonumber \\
\tilde{\rho}_{j,QCD}^1(s)&=&\tilde{\rho}_{\frac{1}{2},\Lambda_b}^1(s)\, ,\,\, \tilde{\rho}_{\frac{3}{2},1,\Lambda_b}^1(s)\, , \, \,\tilde{\rho}_{\frac{3}{2},2,\Lambda_b}^1(s)\, , \, \,\rho_{\frac{1}{2},\Xi_b}^1(s)\, ,\,\, \tilde{\rho}_{\frac{3}{2},1,\Xi_b}^1(s)\, , \, \,\tilde{\rho}_{\frac{3}{2},2,\Xi_b}^1(s)\, ,
\end{eqnarray}

\begin{eqnarray}
\tilde{\rho}_{\frac{1}{2},\Lambda_b}^0(s)&=&\tilde{\rho}_{\frac{1}{2},\Xi_b}^0(s)\mid_{m_s \to 0, \langle\bar{s}s\rangle \to\langle\bar{q}q\rangle, \langle\bar{s}g_s\sigma Gs\rangle \to\langle\bar{q}g_s\sigma Gq\rangle}\nonumber \, , \\
\tilde{\rho}_{\frac{3}{2},1,\Lambda_b}^0(s)&=&\tilde{\rho}_{\frac{3}{2},1,\Xi_b}^0(s)\mid_{m_s \to 0, \langle\bar{s}s\rangle \to\langle\bar{q}q\rangle, \langle\bar{s}g_s\sigma Gs\rangle \to\langle\bar{q}g_s\sigma Gq\rangle}\nonumber \, , \\
\tilde{\rho}_{\frac{3}{2},2,\Lambda_b}^0(s)&=&\tilde{\rho}_{\frac{3}{2},2,\Xi_b}^0(s)\mid_{m_s \to 0, \langle\bar{s}s\rangle \to\langle\bar{q}q\rangle, \langle\bar{s}g_s\sigma Gs\rangle \to\langle\bar{q}g_s\sigma Gq\rangle}\, ,
\end{eqnarray}

\begin{eqnarray}
\tilde{\rho}_{\frac{1}{2},\Lambda_b}^1(s)&=&\tilde{\rho}_{\frac{1}{2},\Xi_b}^1(s)\mid_{m_s \to 0, \langle\bar{s}s\rangle \to\langle\bar{q}q\rangle, \langle\bar{s}g_s\sigma Gs\rangle \to\langle\bar{q}g_s\sigma Gq\rangle}\nonumber \, , \\
\tilde{\rho}_{\frac{3}{2},1,\Lambda_b}^1(s)&=&\tilde{\rho}_{\frac{3}{2},1,\Xi_b}^1(s)\mid_{m_s \to 0, \langle\bar{s}s\rangle \to\langle\bar{q}q\rangle, \langle\bar{s}g_s\sigma Gs\rangle \to\langle\bar{q}g_s\sigma Gq\rangle}\nonumber \, , \\
\tilde{\rho}_{\frac{3}{2},2,\Lambda_b}^1(s)&=&\tilde{\rho}_{\frac{3}{2},2,\Xi_b}^1(s)\mid_{m_s \to 0, \langle\bar{s}s\rangle \to\langle\bar{q}q\rangle, \langle\bar{s}g_s\sigma Gs\rangle \to\langle\bar{q}g_s\sigma Gq\rangle}\, ,
\end{eqnarray}

\begin{eqnarray}
\tilde{\rho}^0_{\frac{1}{2},\Xi_b}(s)&=&\frac{3m_b} {64\pi^2} \langle\frac{\alpha_{s}GG}{\pi}\rangle\int_{x_i}^{1}dx(1-x)(s-\tilde{m}_b^2)+\frac{m_sm_b\langle\bar{s}s\rangle}{64}\langle\frac{\alpha_{s}GG}{\pi}\rangle\delta(s-m_b^2)\nonumber\\
&&+\frac{3m_sm_b\langle\bar{q}g_s\sigma Gq\rangle}{64\pi^2}\int_{x_i}^{1}dx-\frac{m_bs\langle\bar{s}g_s\sigma Gs\rangle\langle\bar{q}g_s\sigma Gq\rangle}{32T^4}\delta(s-m_b^2)  \, ,
\end{eqnarray}

\begin{eqnarray}
\tilde{\rho}^1_{\frac{1}{2},\Xi_b}(s)&=&\frac{3} {64\pi^2}\langle\frac{\alpha_{s}GG}{\pi}\rangle\int_{x_i}^{1}dxx(1-x)(s-\tilde{m}_b^2)+\frac{m_s\langle\bar{s}s\rangle}{64}\langle\frac{\alpha_{s}GG}{\pi}\rangle\delta(s-m_b^2)\nonumber\\
&&-\frac{3m_s\langle\bar{s}g_s\sigma Gs\rangle}{64\pi^2}\int_{x_i}^{1}dxx+\frac{\langle\bar{s}g_s\sigma Gs\rangle\langle\bar{q}g_s\sigma Gq\rangle}{32T^2}\left(1+\frac{s}{T^2}\right)\delta(s-m_b^2)  \, ,
\end{eqnarray}

\begin{eqnarray}
\tilde{\rho}^0_{\frac{3}{2},1,\Xi_b}(s)&=&\frac{m_b} {256\pi^2} \langle\frac{\alpha_{s}GG}{\pi}\rangle\int_{x_i}^{1}dx(1-x)(s-\tilde{m}_b^2)\nonumber\\
&&+\frac{m_sm_b\langle\bar{s}s\rangle}{256}\langle\frac{\alpha_{s}GG}{\pi}\rangle\delta(s-m_b^2)+\frac{3m_sm_b\langle\bar{q}g_s\sigma Gq\rangle}{256\pi^2}\int_{x_i}^{1}dx\nonumber\\
&&-\frac{m_bs\langle\bar{s}g_s\sigma Gs\rangle\langle\bar{q}g_s\sigma Gq\rangle}{128T^4}\delta(s-m_b^2) \, ,
\end{eqnarray}

\begin{eqnarray}
\tilde{\rho}^1_{\frac{3}{2},1,\Xi_b}(s)&=&\frac{1} {256\pi^2} \langle\frac{\alpha_{s}GG}{\pi}\rangle\int_{x_i}^{1}dxx(1-x)(1+2x)(s-\tilde{m}_b^2)
\nonumber\\
&&-\frac{m_s\langle\bar{s}s\rangle}{192}\langle\frac{\alpha_{s}GG}{\pi}\rangle\int_{x_i}^{1}dxx\delta(s-\tilde{m}_b^2)+\frac{m_s\langle\bar{s}s\rangle}{256}\langle\frac{\alpha_{s}GG}{\pi}\rangle\delta(s-m_b^2)\nonumber\\
&&+\frac{m_s\langle\bar{s}g_s\sigma Gs\rangle}{256\pi^2}\int_{x_i}^{1}dxx(1-4x)\nonumber\\
&&-\frac{\langle\bar{s}g_s\sigma Gs\rangle\langle\bar{q}g_s\sigma Gq\rangle}{128T^2}\left(\frac{1}{3}-\frac{s}{T^2}\right)\delta(s-m_b^2) \, ,
\end{eqnarray}

\begin{eqnarray}
\tilde{\rho}^0_{\frac{3}{2},2,\Xi_b}(s)&=&\frac{m_b} {256\pi^2} \langle\frac{\alpha_{s}GG}{\pi}\rangle\int_{x_i}^{1}dx(x-1)(9+4x)(s-\tilde{m}_b^2)
\nonumber\\
&&+\frac{m_sm_b\langle\bar{s}s\rangle}{96}\langle\frac{\alpha_{s}GG}{\pi}\rangle\int_{x_i}^{1}dx\delta(s-\tilde{m}_b^2)-\frac{13m_sm_b\langle\bar{s}s\rangle}{768}\langle\frac{\alpha_{s}GG}{\pi}\rangle\delta(s-m_b^2)\nonumber\\
&&+\frac{m_sm_b\langle\bar{q}g_s\sigma Gq\rangle}{256\pi^2}\int_{x_i}^{1}dx(8x-13)\nonumber\\
&&+\frac{m_b\langle\bar{s}g_s\sigma Gs\rangle\langle\bar{q}g_s\sigma Gq\rangle}{384T^2}\left(\frac{1}{8}+\frac{5s}{T^2}\right)\delta(s-m_b^2) \, ,
\end{eqnarray}

\begin{eqnarray}
\tilde{\rho}^1_{\frac{3}{2},2,\Xi_b}(s)&=&\frac{1}{256\pi^2} \langle\frac{\alpha_{s}GG}{\pi}\rangle\int_{x_i}^{1}dxx(x-1)(11+2x)(s-\tilde{m}_b^2)\nonumber\\
&&+\frac{m_s\langle\bar{s}s\rangle}{192}\langle\frac{\alpha_{s}GG}{\pi}\rangle\int_{x_i}^{1}dxx\delta(s-\tilde{m}_b^2)-\frac{13m_s\langle\bar{s}s\rangle}{768}\langle\frac{\alpha_{s}GG}{\pi}\rangle\delta(s-m_b^2)\nonumber\\
&&+\frac{m_s\langle\bar{q}g_s\sigma Gq\rangle}{256\pi^2}\int_{x_i}^{1}dxx(4-x)\nonumber\\
&&+\frac{\langle\bar{s}g_s\sigma Gs\rangle\langle\bar{q}g_s\sigma Gq\rangle}{384T^2}\left( 1-\frac{s}{3T^2} \right)\delta(s-m_b^2)  \, .
\end{eqnarray}

\section*{Acknowledgements}
This  work is supported by National Natural Science Foundation, Grant Number 12175068, and Postgraduate Students Innovative Capacity Foundation of Hebei Education Department, Grant Number CXZZBS2023146.

\end{document}